\documentclass[]{aastex631}
\usepackage{comment}
\usepackage{xcolor,graphicx,lipsum}
\usepackage[caption=false]{subfig}
\usepackage{ multirow}
\usepackage{enumitem}

\newcommand{\changes}[1]{{\color{red} }}

\begin{document}

\title{Probing the accreting millisecond X-ray pulsar SAX~J1808.4--3658 using the evolution of its spectral and aperiodic timing properties}

\author{Aman Kaushik}
\email{aman.kaushik@tifr.res.in}
\affiliation{Department of Astronomy and Astrophysics, Tata Institute of Fundamental Research, 1 Homi Bhabha Road, Colaba, Mumbai 400005, India} 
\author{Yash Bhargava}
\email{yash.bhargava\_003@tifr.res.in}
\affiliation{Department of Astronomy and Astrophysics, Tata Institute of Fundamental Research, 1 Homi Bhabha Road, Colaba, Mumbai 400005, India}%
\author{Sudip Bhattacharyya}
\email{sudip@tifr.res.in}
\affiliation{Department of Astronomy and Astrophysics, Tata Institute of Fundamental Research, 1 Homi Bhabha Road, Colaba, Mumbai 400005, India}%
\author{Maurizio Falanga}
\altaffiliation{Deceased: 2025 March 6}
\affiliation{International Space Science Institute (ISSI), Hallerstrasse 6, 3012 Bern, Switzerland}
\affiliation{Physikalisches Institut, University of Bern, Sidlerstrasse 5, 3012 Bern, Switzerland}

\begin{abstract}

Understanding accretion components in neutron star (NS) low-mass X-ray binary (LMXB) systems is important to probe fundamental aspects of accretion mechanism and evolution of the system constraining its physical properties. Here, we present  spectral and aperiodic timing analyses of the \textit{NICER} and \textit{AstroSat} data from the accretion-powered millisecond X-ray pulsar (AMXP) SAX J1808.4$-$3658 during its 2022 outburst.
We find that emissions from a softer accretion disk and a harder, centrally located, compact, partially covering, Comptonizing corona explains the continuum spectra from the source throughout the outburst. 
The disk inner edge temperature, the coronal electron temperature and photon index are found to be around $\sim 0.5-0.9$~keV, a few keV and $\sim 1.1-1.8$, respectively, during the entire outburst.
We also find an intrinsic atomic hydrogen medium in the system, which substantially and systematically
evolved throughout the outburst.
We detect two broadband aperiodic features ($\sim 0.004-2$~Hz; $\sim 10-100$~Hz), with the former having a  significant hard lag of $\sim 11$\,ms between $1.5-10.0$~keV and $0.5–1.5$~keV photons.
We conclude that both the disk photons and the photons up-scattered by
the corona contributed to each aperiodic feature, with the disk and the corona contributing more to the low and high-frequency ones, respectively. 

\end{abstract}

\section{Introduction} \label{sec:intro}

Accretion-powered millisecond X-ray pulsars \citep[AMXPs; ][]{campana2018accreting, patruno2020accreting, salvo2021accretion} are rapidly spinning neutron stars (NSs) in low-mass X-ray binary (LMXB) systems. 
In an NS LMXB, the NS accretes matter via an accretion disk from a relatively low-mass ($\lesssim 1$~M$_\odot$) companion star as the latter fills its Roche lobe.
This matter close to the NS primarily emits X-rays, which can be useful to probe some extreme aspects of physics and astronomy, such as
the strong gravity and dense matter of NSs \citep{psaltis2008probes, 2010AdSpR..45..949B}
and the accretion processes in the strong gravity region, that cannot be studied by doing experiments in laboratories.
The large specific angular momentum of the accreted matter can spin up the NS to frequencies of a few hundred Hz \citep{1991PhR...203....1B}.
In the case of AMXPs, which are a subset of NS LMXBs, the NS magnetic field is strong enough \citep[typically, $\sim 10^8-10^9$~G; e.g., ][]{2015MNRAS.452.3994M} to truncate the accretion disk and channel the matter onto it's magnetic poles. 
The resulting hot spots on the spinning NS cause the observed X-ray intensity to vary with the stellar spin frequency $\nu$. 
Hence such systems are called AMXPs, as they are X-ray pulsars powered by accretion and having periods of milliseconds. These AMXPs, which are transient NS LMXB, accrete through a series of outburst and quiescent phases.
During an outburst, which lasts for weeks, the accretion rate increases by orders of magnitude.
Then as the accretion decays, the source goes into a quiescent period which can range from a few years to decades, depending on the system \citep{2025ApJS..279...57H}.
The source spectral and timing properties can be meaningfully studied only during the outbursts, when at the peak 
the X-ray luminosity typically reaches around a few times $10^{36}$\,erg\,s$^{-1}$ \citep{gilfanov1998millisecond,2004ApJ...614L..49C}.

In order to utilize the  accretion to study the extreme physics of NSs, it is essential to identify and probe various accretion components. 
Fitting of spectral data with physical models can contribute to such understanding. 
For an  NS LMXB,
typically one expects emissions from an accretion disk, the NS surface and a corona (hot plasma) covering the disk and/or the NS, fully or partially \citep{2011MNRAS.411.2717M}. 
The disk and NS emissions could be multicolor blackbody and single temperature blackbody, respectively.
The hot electrons of the corona Comptonize or upscatter the seed photons from the disk and/or the NS.
Indeed, previous studies have shown that the broadband X-ray spectra of AMXPs could be well fit with the emission from a multicolor blackbody disk (e.g. see \citealt{2009MNRAS.396L..51P, 2011MNRAS.417.1454K, 2023MNRAS.519.3811S, 2024MNRAS.532.3961P}) and/or the NS surface (e.g. see \citealt{2010MNRAS.407.2575P, 2017MNRAS.466.2910S}), 
and also from a Comptonizing thermal hot plasma or corona 
\citep[e.g.,][]{falanga2005integral, 2010MNRAS.407.2575P, 2012A&A...545A..26F, 2017MNRAS.466.2910S, 2021A&A...649A..76L, 2023ApJ...958..177L, 2023MNRAS.519.3811S, 2024MNRAS.532.3961P}.

The accretion flow and components can also be probed by studying various timing features of NS LMXBs, including AMXPs \citep{1989ARA&A..27..517V}. 
The aperiodic X-ray variability components can be broadly grouped into high-frequency phenomena (including kilo-Hertz and hecto-Hertz quasi periodic oscillations (QPOs)) and a low-frequency complex ($0.01-100$~Hz). 
The latter could include multiple broadband, peaked-noise and QPO components 
(e.g., see \citealt{2004NuPhS.132..664V, 2007AIPC..924..629L, 2011ApJ...729....9K, 2015ApJ...806...90B}).
According to \citet{1997MNRAS.292..679L}, a low-frequency broadband or red noise could originate from the variability of disk parameters at different radii which may cause variation of the accretion rate and hence the observed flux at different time scales.
The fluctuations originate at larger radii in the disk and propagate inwards causing the high-frequency variability in the inner parts of the accretion flow
\citep{2006MNRAS.367..801A, 2020MNRAS.498.2757G, 2024ApJ...975..190T, 2024MNRAS.tmp.2438U}.

In this paper, we report the analysis of observations from {Neutron star Interior Composition Explorer} {(\it NICER)} and {\it AstroSat} of the AMXP SAX~J1808.4$-$3658 (hereafter SAX~J1808). It was discovered in 1996 with the {\it BeppoSAX} satellite \citep{1998A&A...331L..25I}. 
During another outburst in 1998, the NS spin frequency (401 Hz) was measured from X-ray pulsations \citep{1998Natur.394..344W}.
The binary orbital period of SAX~J1808 is $2.01$ hr  \citep{1998Natur.394..346C}, and the companion star may be a brown dwarf with an estimated mass of $\approx 0.05$ M$_\odot$ \citep{2001ApJ...557..292B}. SAX~J1808 was the first (and so far the only) AMXP where optical millisecond pulsations were observed  \citep{ambrosino2021optical}.

SAX~J1808, being a transient with outbursts observed every 2--4 years (see \citealt{illiano2023timing} and references therein and \citealt{2025ATel17323....1R}), is an ideal source to study the evolution of spectral and timing aspects across a large range of accretion rates \citep{1999A&A...343L..45E, poutanen2003nature}. 
This can provide an excellent opportunity to probe the accretion processes for wide ranges of parameter values. 
The X-ray spectra of SAX~J1808 was modeled with a thermal emission from an accretion disk \citep{2009ApJ...694L..21C,2009MNRAS.396L..51P, 2011MNRAS.417.1454K, 2023MNRAS.519.3811S} or the NS surface \citep{2002MNRAS.331..141G, 2009A&A...493L..39P}, a Comptonized emission and reprocessed emission line components \citep{2023MNRAS.519.3811S}. 
The aperiodic timing analysis showed  kilohertz QPOs (e.g., 410\,Hz)  
and low-frequency broadband noise \citep[e.g.,][]{2001MNRAS.323L..26U, 2004NuPhS.132..664V, 2015ApJ...806...90B}.

The 2022 outburst of SAX~J1808 was detected by MAXI/GSC nova alert system on 19 August 2022 \citep{2022ATel15563....1I}. 
The coherent timing analysis of the pulsations observed in this outburst has already been studied by \citet{2023ApJ...942L..40I} showing hints of orbital contraction, in contrast to the previously observed orbital expansion.
In this paper, we primarily aim to study the evolution of spectral and aperiodic timing properties throughout the 2022 outburst of SAX~J1808.
For this purpose, we mainly use the {\it NICER} data, as {\it NICER} observed the source throughout the outburst. However, we also report the analysis of the contemporaneous {\it AstroSat} data.
We describe the data and instruments in section~\ref{sec:obs_analysis}, and detail the spectral analysis and results in section~\ref{sec:spectral}.
Results from the aperiodic timing analysis are presented in section~\ref{sec:aperiodic}.
We discuss the implications of our results in section~\ref{sec:discussion} and summarize our work in section~\ref{sec:summary}.

\clearpage
\section{Observation and Data analysis} \label{sec:obs_analysis}

\subsection{NICER} \label{subsec:nicer}

 \textit{NICER} \citep[][]{gendreau2012neutron}, launched in 2017, is an International Space Station (ISS) payload devoted to the study of NS through soft X-ray timing and spectroscopy. \textit{NICER} has facilitated rotation-resolved spectroscopy of emissions from NSs within the soft X-ray band (0.2--12 keV) with exceptional sensitivity.
The X-ray Timing Instrument (XTI) is the primary science instrument of \textit{NICER} \citep{Gendreau2016SPIE.9905E..1HG} which consists of an array of 56 X-ray photon detectors. These detectors measure both the energy and the arrival time of the incoming X-ray photons. The precise timing and positioning measurements of the instrument is maintained by a Global Positioning System (GPS) receiver on board the system. The X-ray photons can be time-tagged with an accuracy of less than 300 nanoseconds. 
SAX~J1808 was monitored with \textit{NICER} from 2022 August 19 (MJD 59810) until 2022 October 31 (MJD 59883).  
For our analysis, we consider 17 observations from MJD 59810 to 59825 (ObsID: from 5050260101 to 5050260106 and from 5574010101 to 5574010111), as the intensity of the source became very low and the data have large gaps between the observations after that. We list the details of the observations analyzed in this work in Table~\ref{tab:NICER_TABLE}. To process the data and generate event file for each observation, we use the standard pipeline processing tool for \textit{NICER} data (\texttt{nicerl2}), included in the \textsc{heasoft} v6.33 software package, along with all standard \textit{NICER} analysis tools (as part of \textsc{nicerdas 2024-02-09\_v012}). 
For generating light curves and spectra for each observation, we use the \texttt{nicerl3-lc} and \texttt{nicerl3-spect} pipelines respectively with the background model \texttt{SCORPEON v23  using \texttt{bkgformat=file}} \citep{SCORPEON2024HEAD...2110536M}.  To make sure that the light curve and hardness intensity diagram (HID) of each observation is not contaminated by non-X-ray flares,
we run the \texttt{nicer-l2} command with \texttt{cor\_range}=$``1.5-^{*}"$. Since, precipitation flares are almost always limited to \texttt{cor\_sax} $< 1.5$, we take this value for all the ObsIDs except for observations 5574010104 and 5574010108, where a higher threshold for \texttt{cor\_sax} of 2.5 and 3.0 respectively is required to filter out the electron flares. After filtering out the non-source flaring events, we re-extract the light curves and spectral products using the tools mentioned above for further analysis. We have not detected any type--I X-ray burst in the data. For our spectral analysis, we have limited the energy range to 0.5--10.0\,keV.

\subsection{AstroSat/LAXPC}

\textit{AstroSat}, launched in 2015, is India's first dedicated multi-wavelength astronomy satellite \citep{2006AdSpR..38.2989A, 2014SPIE.9144E..1SS}. It is equipped with four main scientific pointing payloads: (i) the Large Area X-ray Proportional Counters \citep[LAXPC; ][]{2016SPIE.9905E..1DY,Yadav2016SPIE.9905E..1DY,Yadav2017CSci..113..591Y,2017ApJS..231...10A}, (ii) the Soft X-ray Telescope \citep[SXT; ][]{2016SPIE.9905E..1ES}, (iii) the Cadmium-Zinc-Telluride Imager  \citep[CZTI; ][]{2016SPIE.9905E..1GV}, and (iv) the Ultra-Violet Imaging Telescope \citep[UVIT; ][]{2016SPIE.9905E..1FS}. 
For our analysis of SAX~J1808, we  focus on the data observed with \textit{AstroSat}/LAXPC from 26\,August\,2022 to 27\,August\,2022  (Table~\ref{tab:NICER_TABLE}). 

\textit{AstroSat}/LAXPC has three identical co-aligned proportional counters operating in the 3–80 keV energy range named LAXPC10, LAXPC20, and LAXPC30. Each \textit{AstroSat}/LAXPC detector consists of five layers and independently records the arrival time of each photon with a time resolution of 10\,$\mu$s and deadtime of 42\,$\mathrm{\mu}$s. The orbit-wise level 1 data is downloaded from AstroBrowse\footnote{\url{https://astrobrowse.issdc.gov.in/astro_archive/archive/Home.jsp}} with observation ID: T05$\_$048T01$\_$9000005318. 
For our analysis, we use only the LAXPC20 detector as LAXPC10 has been operating at low gain since 2018 and detector LAXPC30\footnote{\url{http://astrosat-ssc.iucaa.in/}} had been switched off early in the mission due to a leak in the detector. 
To minimize the background, only the top layer (L1) data are used, which restricts the operational energy range to 3--20~keV.
We reduce the \textit{AstroSat}/LAXPC data using the pipeline \texttt{LAXPCSOFTWARE22AUG15}\footnote{\url{http://astrosat-ssc.iucaa.in/uploads/laxpc/LAXPCsoftware22Aug15.zip}} \citep{Antia2021JApA...42...32A, Misra2021JApA...42...55M}. The software includes
tools to reduce the level 1 data to level 2 data, calibration files, responses, etc. We obtain the level 2 data and standard good time intervals (GTIs) using the tools from the pipeline. We also extract the energy-dependent spectra, light curves, background spectra, and background light curves using the tools available in \texttt{LAXPCSOFTWARE22AUG15}. We have not detected any type–I X-ray burst in the data.

\begin{table*}
    \centering
\caption{\textit{NICER} and \textit{AstroSat}/LAXPC observation log table for 
the source SAX~J1808 with net exposure times  
(section~\ref{sec:obs_analysis}).
}     \label{tab:NICER_TABLE}
    \begin{tabular}{|c|c|c|c|c|r|}
    \hline
    Instrument & Obs\ ID & Obs\ no. & Start time\ (MJD, UTC) & Stop time\ (MJD, UTC) & Exposure\ (ks)\\
    
    \hline
     \multirow{17}{*}{\textit{NICER}} & 5050260101&1& 59810.5916 & 59810.9423 & 4.0 \\
    & 5050260102& 2 & 59810.9790 & 59811.9555 & 10.1 \\
    & 5050260103& 3 & 59812.1401 & 59812.9424 & 7.3 \\
    & 5050260104& 4 & 59812.9877 & 59813.9587& 9.5 \\
    & 5050260105& 5 & 59814.0188 & 59814.8775& 10.8 \\
    & 5574010101& 6 &59814.9221 & 59814.9421 & 0.6 \\
    & 5574010102& 7 &59815.1155 & 59815.8426 & 5.9 \\
    & 5050260106& 8 &59815.5024 & 59815.6484 & 1.9 \\
    & 5574010103& 9 &59816.0829 & 59816.9419 & 13.2\\
    & 5574010104& 10 &59817.0676 & 59817.9743 & 8.9 \\
    & 5574010105& 11 &59818.0343 & 59818.9888& 10.1 \\
    & 5574010106& 12 &59819.0453 & 59819.9743 & 9.3 \\
    & 5574010107& 13 &59820.0220 & 59820.9421 & 6.4 \\
    & 5574010108& 14 &59821.1331 & 59821.9745 & 7.1 \\
    & 5574010109& 15 &59822.0154 & 59822.9896 & 10.5 \\
    & 5574010110& 16 &59823.0005 & 59823.9703 & 10.6 \\
    & 5574010111& 17 &59824.0342 & 59824.9888 & 6.5 \\
    \hline
    \textit{AstroSat}/LAXPC & 9000003158 & & 59817.30055 & 59818.57988 & 26.9 \\
    \hline
    \end{tabular}
\end{table*}

\section{Spectral results} \label{sec:spectral}

\subsection{Hardness Intensity Diagram} \label{subsec:hid}

\begin{figure*}
    \centering
\includegraphics[width=0.49\columnwidth]{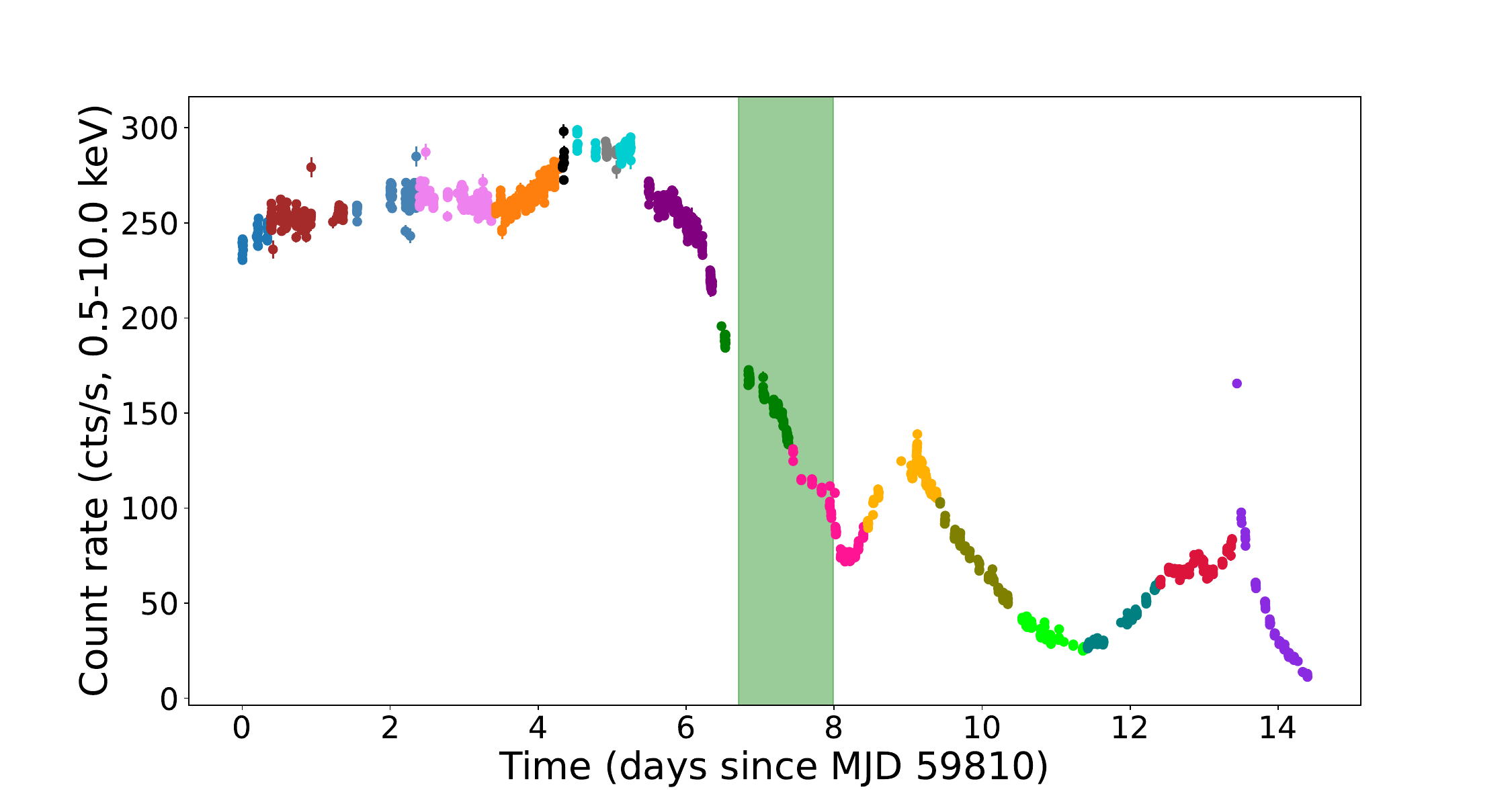}
\includegraphics[width=0.49\columnwidth]{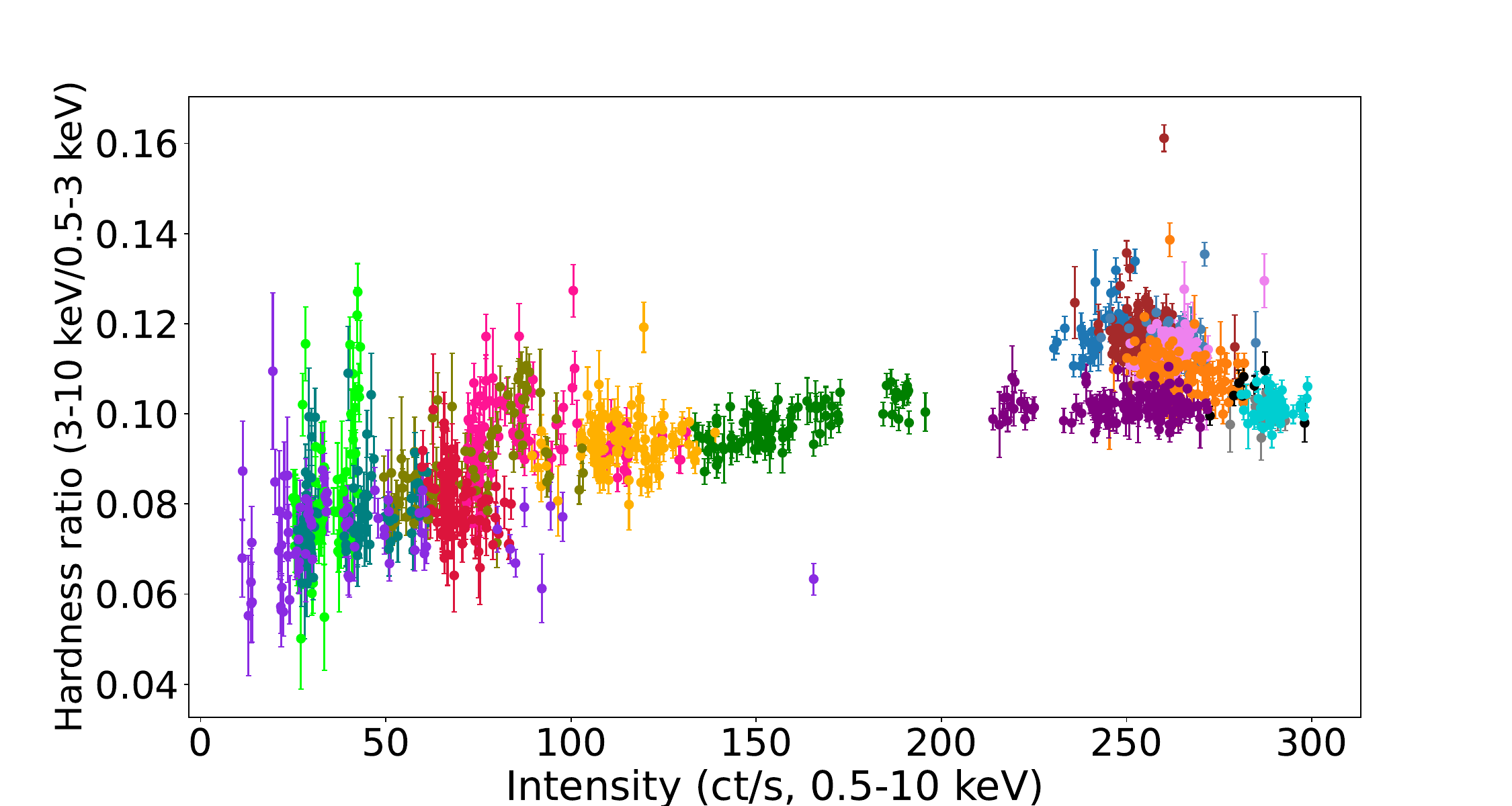}
\caption{ 
{\it Left panel:} The 0.5--10.0\,keV \textit{NICER} light curve for the observation IDs 5050260101 to 5050260106 and from 5574010101 to 5574010111 of the source SAX~J1808. The green  band shows the time range of \textit{AstroSat} observations. 
{\it Right panel:} The hardness-intensity diagram (HID) of \textit{NICER} observations, with hardness defined in the ranges 3--10\,keV/0.5--3\,keV. The total photon counts are taken from the energy range 0.5--10\,keV. The color scheme in both the panels are same and 1$\sigma$ error bar has been reported (see sections~\ref{subsec:nicer},~\ref{subsec:hid}).}
    \label{fig:lightcurve_hid}
\end{figure*}

We extract the 100\,s binned light curve for all the observations in the 0.5--10.0\,keV energy range for \textit{NICER} using \texttt{nicerl3-lc} and depict the evolution of the source in Figure~\ref{fig:lightcurve_hid} left panel. The count rate decreases from $\sim{300}$ counts/sec to $\sim{11}$ counts/sec as the outburst progresses. To investigate if the change in count rate is associated with a change in the hardness, we plot the hardness-intensity diagram (HID; right panel of Figure~\ref{fig:lightcurve_hid}). 
The soft band and the hard band for the HID are 0.5--3.0\,keV and 3.0--10.0\,keV, respectively. 
These energy ranges are motivated by the spectral decomposition of the source displayed in Figure~\ref{fig:102_spectrafit}, showing that the soft band (0.5--3.0\,keV) is dominated by the accretion disk emission (\texttt{diskbb}) and the hard band (3.0--10.0\,keV) is dominated by a Comptonized emission (see section~\ref{subsec:preliminary} for more details). We do not observe significant variation in the HID and the hardness remains low as the source evolves. Hence, we conclude that the source remain in the soft state throughout the outburst. 

\subsection{Broadband spectral analysis} \label{subsec:preliminary}

Here, we briefly describe the spectral fitting for data of the observation ID 5050260102, as it has high count rate and has one of the longest exposure time among all the observations but we have applied similar modeling to all the \textit{NICER} spectra. For spectral fitting, we use the software \texttt{XSPEC v12.14.0h} and use $\chi^{2}$ statistics. 
We report the 1$\sigma$ uncertainty in the parameters for all the observations.
To account for the absorption by interstellar and intrinsic neutral hydrogen medium,  we use the \texttt{XSPEC} model \texttt{tbabs}. 
We consider abundances from \citet{2000ApJ...542..914W} and the cross-sections from \citet{1996ApJ...465..487V} to model the spectrum. 
We limit the spectrum for each observation to an energy range where the source counts are above the background counts (e.g. for observation 5050260102, the considered energy range is 0.5--10.0\,keV).

\begin{figure*}
    \centering
\includegraphics[width=0.49\columnwidth]{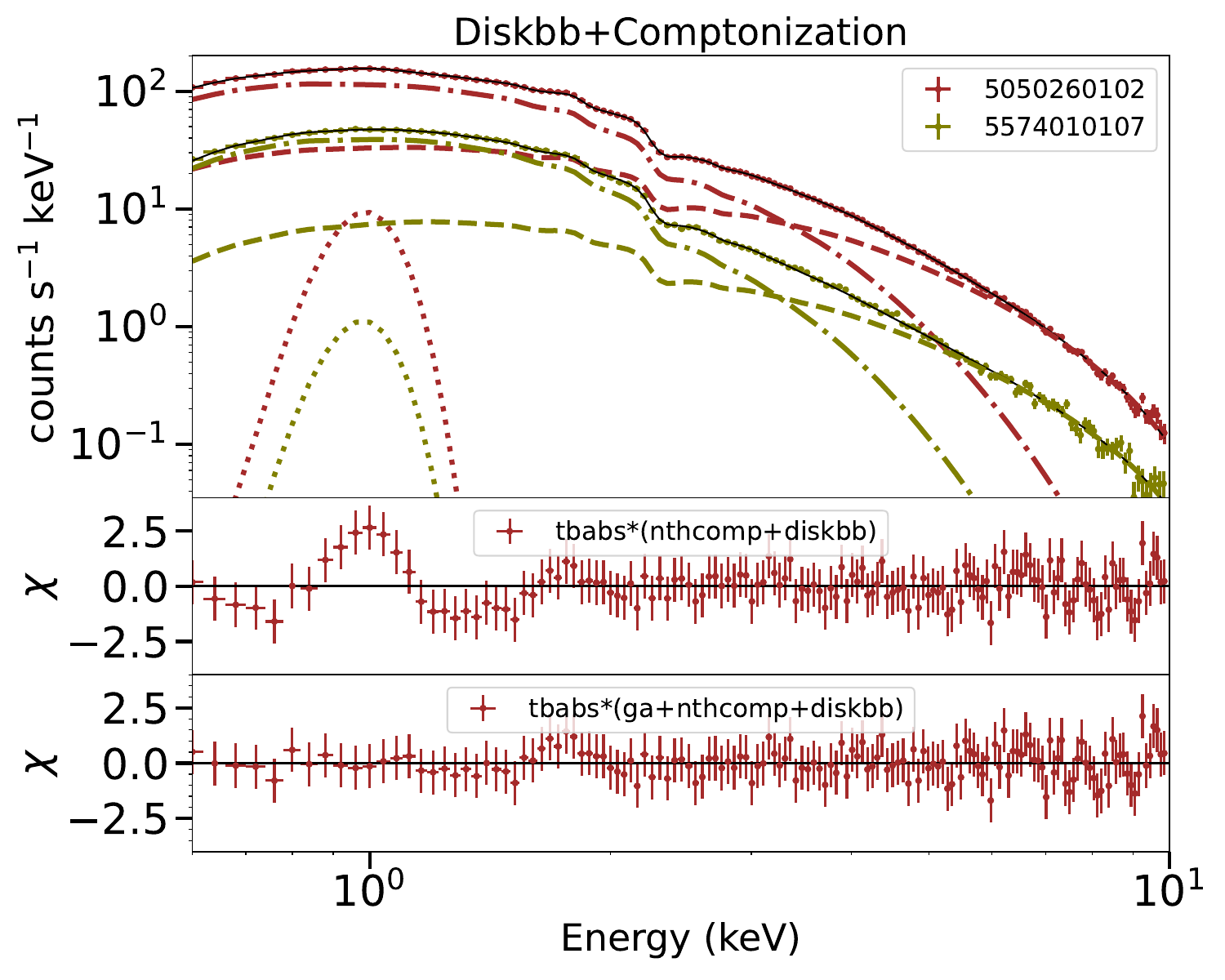}
\includegraphics[width=0.49\columnwidth]{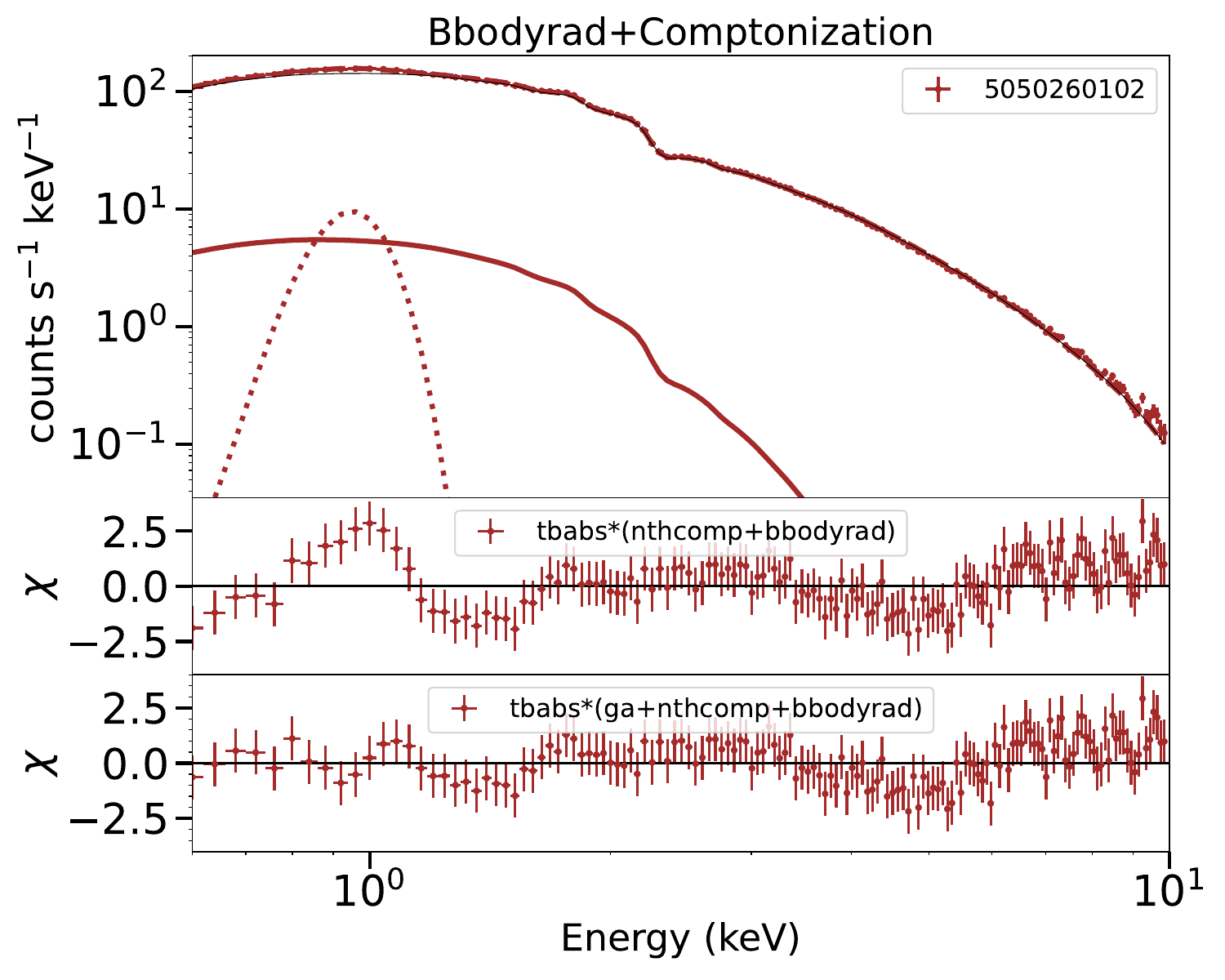}
\caption{Spectral fitting of the \textit{NICER} data from the source SAX~J1808 (see section~\ref{subsec:preliminary}). 
{\it Left plot:} The unabsorbed best-fit spectral model (solid line) for the observations 5050260102 (red: as depicted in the lightcurve) and 5574010107 (green: as depicted in the lightcurve) using the \texttt{XSPEC} model \texttt{tbabs*(gaussian+nthcomp+diskbb}) with a $\chi^{2}$/dof of 107/142 for the former. 
For both the observations, the dash-dot curve shows the disk spectrum, the dashed line shows the Comptonized spectrum, and the dotted curve is for a Gaussian line. 
{\it Right plot:}  The spectral fitting of the observation 5050260102 using the \texttt{XSPEC} model \texttt{tbabs*(gaussian+nthcomp+bbodyrad}) with a $\chi^{2}$/dof of 238/144. 
Data and model spectrum are shown in this plot.
Here, the solid (red), dashed and dotted lines show the blackbody, Comptonized, and Gaussian components of the spectrum, respectively.
The two bottom panels in each plot display the fit residuals without the Gaussian line (middle panel) and with the Gaussian line (bottom panel).
This figure shows that a disk component fits the lower energy part of the continuum spectrum much better than a blackbody component.
Moreover, the intersection points of the dash-dot and dashed curves of the left plot show that the spectrum is dominated by the disk emission below $\sim3$\,keV and by the Comptonized emission above $\sim3$\,keV. 1$\sigma$ error bar has been reported in the middle and bottom residual panels in both the plots (section~\ref{subsec:preliminary}).}
    \label{fig:102_spectrafit}
\end{figure*}

We take the \texttt{XSPEC} model \texttt{tbabs*(diskbb)} and include a Comptonized emission model, \texttt{nthcomp}  \citep{10.1093/mnras/283.1.193, 1999MNRAS.309..561Z} resulting in $\chi^{2}$/degrees of freedom (dof) = 146/144. 
The \texttt{XSPEC} model component \texttt{nthcomp} involves a seed photon source as a blackbody  or a disk-like multi-color blackbody and 
the seed photons are up-scattered by hot electrons resulting in the output spectrum. We assume that the source of the seed photons is the accretion disk and thus we tie the seed temperature k$T_\mathrm{bb}$ to the disc inner edge temperature $\mathrm{k}T_\mathrm{in}$ \citep[e.g.][]{2023ApJ...955..102B, 2024MNRAS.532.4486A}. 
We find a residual at $\sim 1$~keV, which was also seen in the \textit{AstroSat} observation of SAX~J1808.4--3658 during its 2019 outburst  \citep{2023MNRAS.519.3811S} and has been detected in other sources (e.g. Hercules X-1 \citealt{2022ApJ...936..185K}, Cygnus X-2 \citealt{2022ApJ...927..112L} and see   \citealt{2024arXiv240702360C} for other examples). It is a strong emission-line complex centered on $\sim 1$~keV and is a blend of multiple atomic transition line emissions, like Ne Ly$\alpha$, Ne Ly$\beta$, Ne He-like-$\alpha$ \& $\beta$, etc., the strongest among which is observed from Ne Ly$\alpha$ at $\sim$1.008\,keV \citep{1995ApJ...449L..41A}. In some sources a broader feature has been observed ($0.5-2.0$~keV) which also includes emissions from H-like Ne, H-like Mg and He-like Mg, O Ly$\alpha$ and O He$\alpha$ \citep{2024arXiv240702360C}.
We model it using a Gaussian with the fixed line-width of 0.08\,keV (total model \texttt{tbabs*(gaussian+nthcomp+diskbb)}), reducing the $\chi^{2}$/dof to 107/142 with a F-test score of $\sim27$ and probability of improvement by chance $\sim1.5\times10^{-10}$. In support of its significance, we find that in all spectra the normalization deviate from the continuum by more than $\sim3\sigma$.
We note that the model \texttt{tbabs*(diskbb+nthcomp)} is statistically sufficient to describe the spectrum but we still include the Gaussian component as 
(i) it has been seen in earlier spectral analysis of the source and (ii) a structured residual is clearly visible in Figure~\ref{fig:102_spectrafit}.
For completeness, we also use the model \texttt{tbabs*(gaussian+nthcomp+bbodyrad}) \citep{2011MNRAS.410.1513W, 2019MNRAS.483..767D}. 
For instance, for the observation 5050260102 this model results in  $\chi^{2}$/dof = 238/144 (right panel of Figure~\ref{fig:102_spectrafit}). 
Additionally, we note that the fit parameter values of the electron temperature k$T_\mathrm{e}$ and the photon index $\Gamma$ of the Comptonizing component are not physically acceptable.
On the other hand, the fit statistics for the model \texttt{tbabs*(gaussian+nthcomp+diskbb}) for all the spectra are sufficiently good. 
Nevertheless, we do not consider this as our final model, because while calculating the $1\sigma$ uncertainty, k$T_\mathrm{e}$ and $\Gamma$ are unconstrained.
This means $\Gamma$ approaches the value 1, implying  infinite optical depth of the Comptonizing medium, and k$T_\mathrm{e}$ gets pegged to the highest value of the temperature range considered much beyond the operational range of \textit{NICER} or \textit{AstroSat}/LAXPC and therefore suggesting that a better model description may be needed.

\begin{figure}
    \centering
\includegraphics[width=0.5\columnwidth]{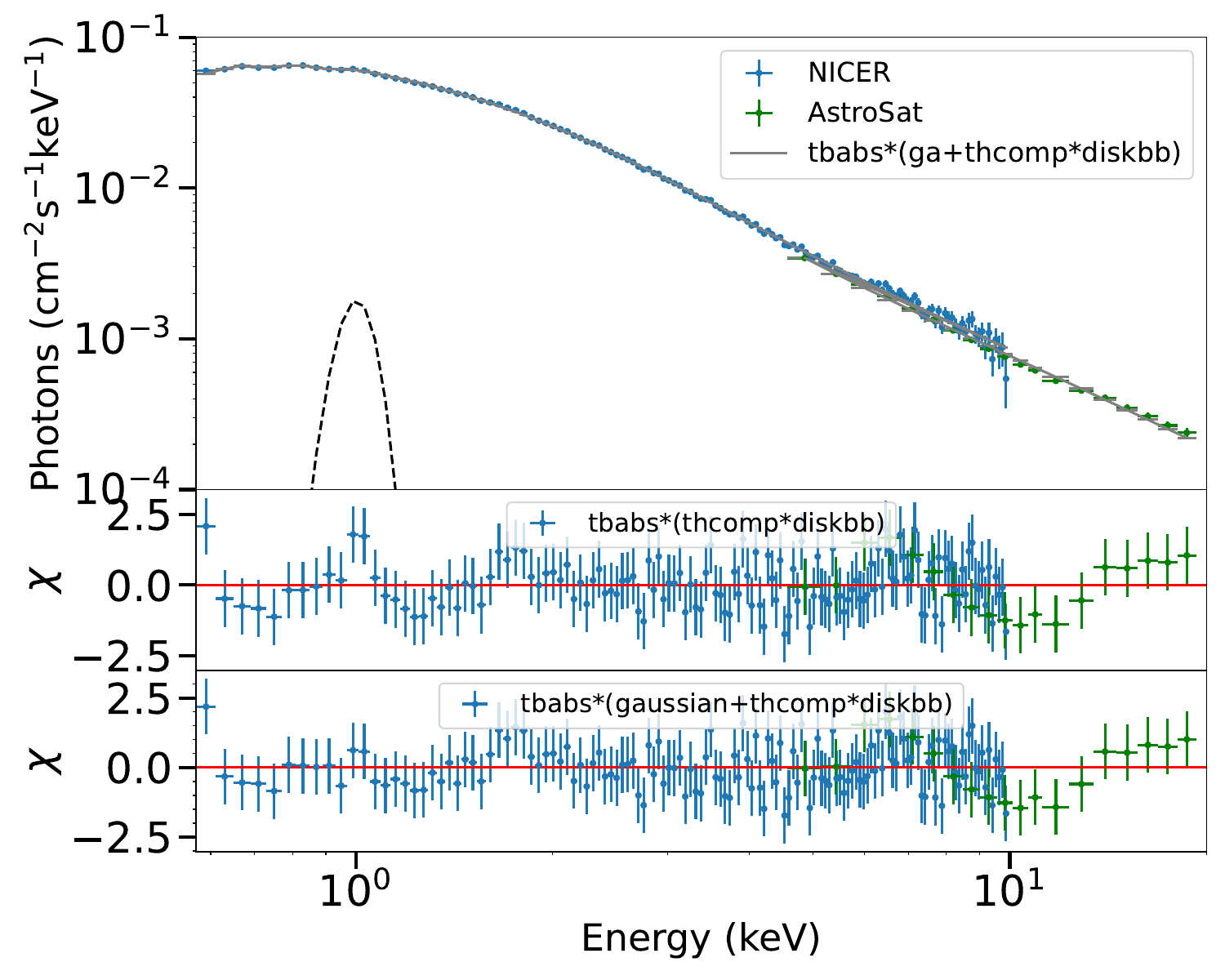}
\caption{This figure shows the spectral fitting (top panel) for the \textit{NICER} (blue) and \textit{AstroSat}/LAXPC (green) contemporaneous data  from the 2022 outburst of the source SAX~J1808 using the \texttt{XSPEC} model \texttt{tbabs*(gaussian+thcomp$\otimes$diskbb)} in the energy range 0.5--20.0\,keV ($\chi^{2}$/dof = 113/151). 
The middle panel shows the residual plot ($\chi$ = (Data$-$Model)/{error}) without a Gaussian component. The bottom panel shows the residual plot after including a Gaussian component at $\sim 1$\,keV. 1$\sigma$ error bar has been reported in all three panels (section~\ref{subsec: spectral evolution}).}
\label{fig:spectralfit}
\end{figure}

To solve the unconstrained k$T_\mathrm{e}$ and a low $\Gamma$, we replace \texttt{nthcomp} with \texttt{thcomp} \citep{2020MNRAS.492.5234Z}  in the \texttt{XSPEC} model \texttt{tbabs*(gaussian+thcomp$\otimes$diskbb)}. 
While the use of the \texttt{nthcomp} component is suitable to estimate the seed photon flux and the Comptonized flux separately, 
the convolution model \texttt{thcomp} is an updated version of \texttt{nthcomp} and 
agrees with the actual Monte Carlo spectra from Comptonization much better than \texttt{nthcomp} \citep{2020MNRAS.492.5234Z}. 
The component \texttt{thcomp} involves upscattering fraction (\textit{f}) of photons from any seed emission, and gives the low-energy cut-off corresponding to the soft tail of the seed photon spectrum and the high energy rollover corresponding to k$T_\mathrm{e}$. 
Thus, the seed photon source parameters can be simultaneously constrained in such a description (while \texttt{nthcomp} allows for constraint on the seed photon temperature for only two cases, a single temperature blackbody or a multicolor blackbody). For an accurate computation of the convolution model, we use the energy range from 0.01--100\,keV having 1000 logarithmic bins which is significantly larger than the range covered by the instrument response. 
In order to test if the seed photons could have been supplied by a blackbody component, we tried the model \texttt{tbabs*(gaussian+thcomp$\otimes$bbodyrad)} for some of the high flux data. 
The fit is unacceptable in every case, e.g., for the observation 5050260102, the $\chi^{2}/$dof is 238/143, with the covering fraction value tending to 1 (implying the upscattering of all the seed photons) and an unphysically low $n_\mathrm{H}$ value of $\sim10^{-18}$\,cm$^{-2}$.
This shows that a single temperature blackbody is not suitable, and we need the model \texttt{tbabs*(gaussian+thcomp$\otimes$diskbb)} which satisfactorily fits the spectra (see Figure~\ref{fig:spectralfit}). 
We additionally explore if the reflection component (e.g. \texttt{relxillNS}) can jointly fit with the blackbody component. We obtain a similar test statistic (as model \texttt{tbabs*(gaussian+thcomp$\otimes$diskbb)}) for a fewer degrees of freedom, suggesting that the reflection component is also consistent with the data. Noting that Figure~\ref{fig:102_spectrafit} doesn't show any significant residuals at $\sim6.4$\,keV and the fact that inclusion of a reflection component reduces the degrees of freedom without changing the test statistic, we proceed with a simpler model without a reflection component.

\begin{figure*}
    \centering
\includegraphics[width=0.49\columnwidth]{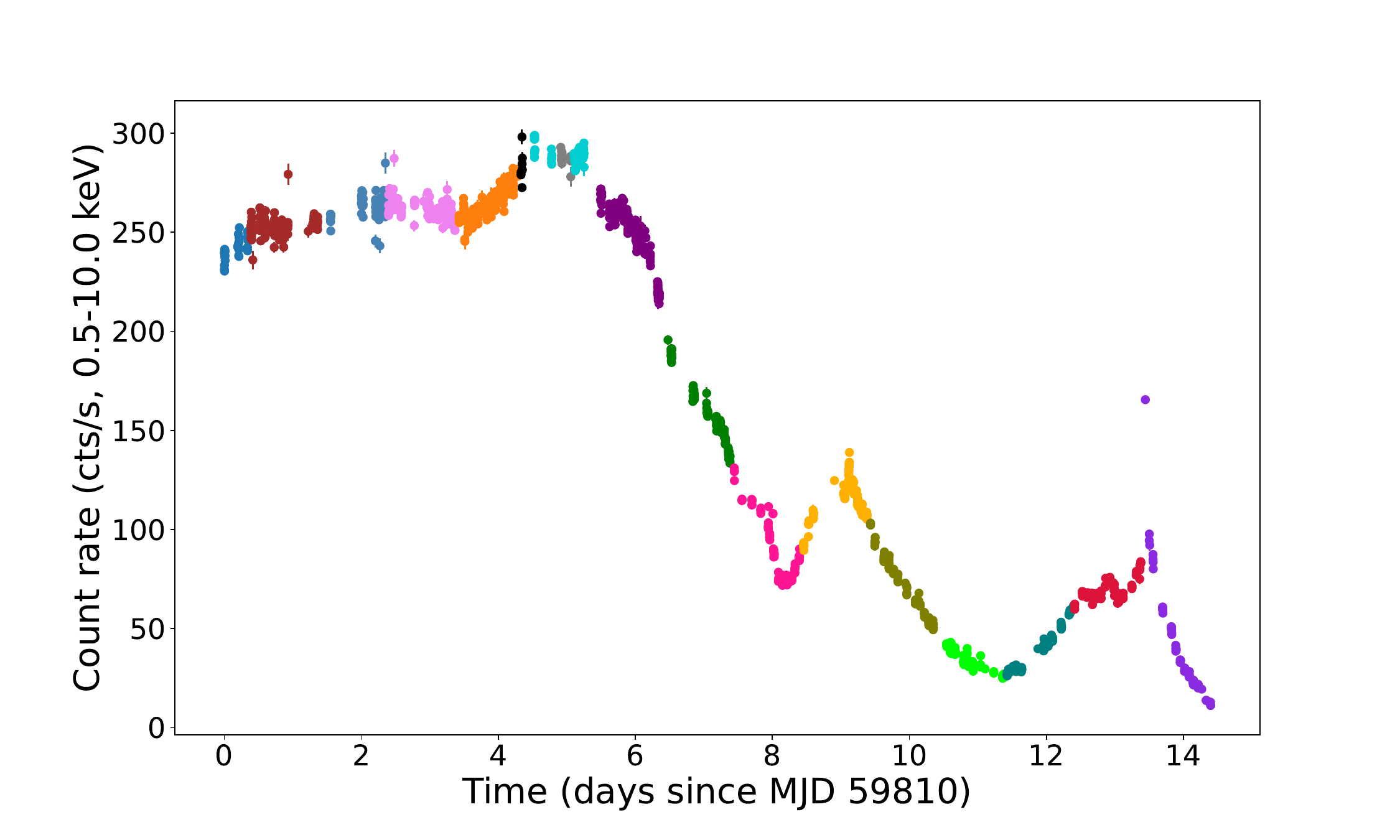}
\includegraphics[width=0.49\columnwidth]{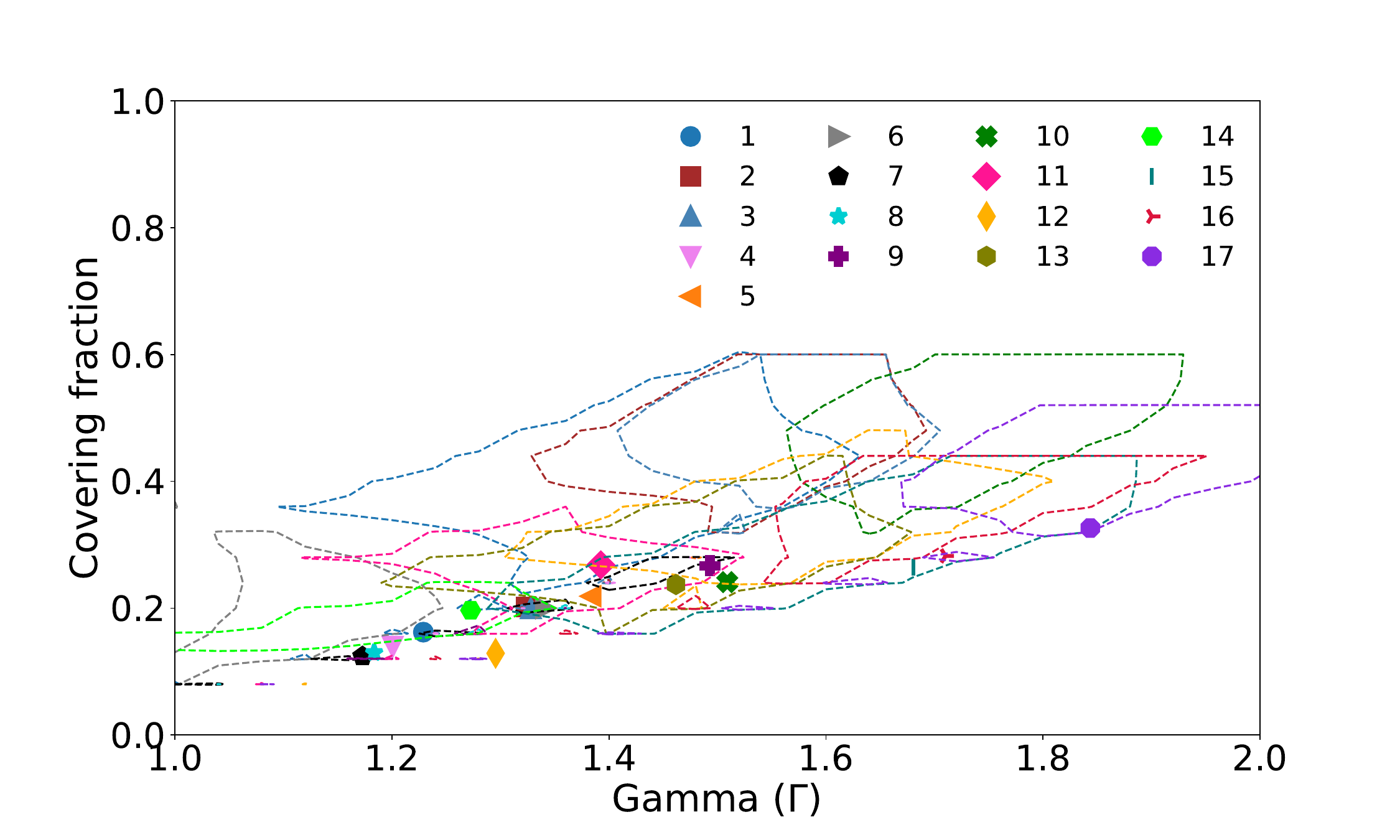}
\includegraphics[width=0.49\columnwidth]{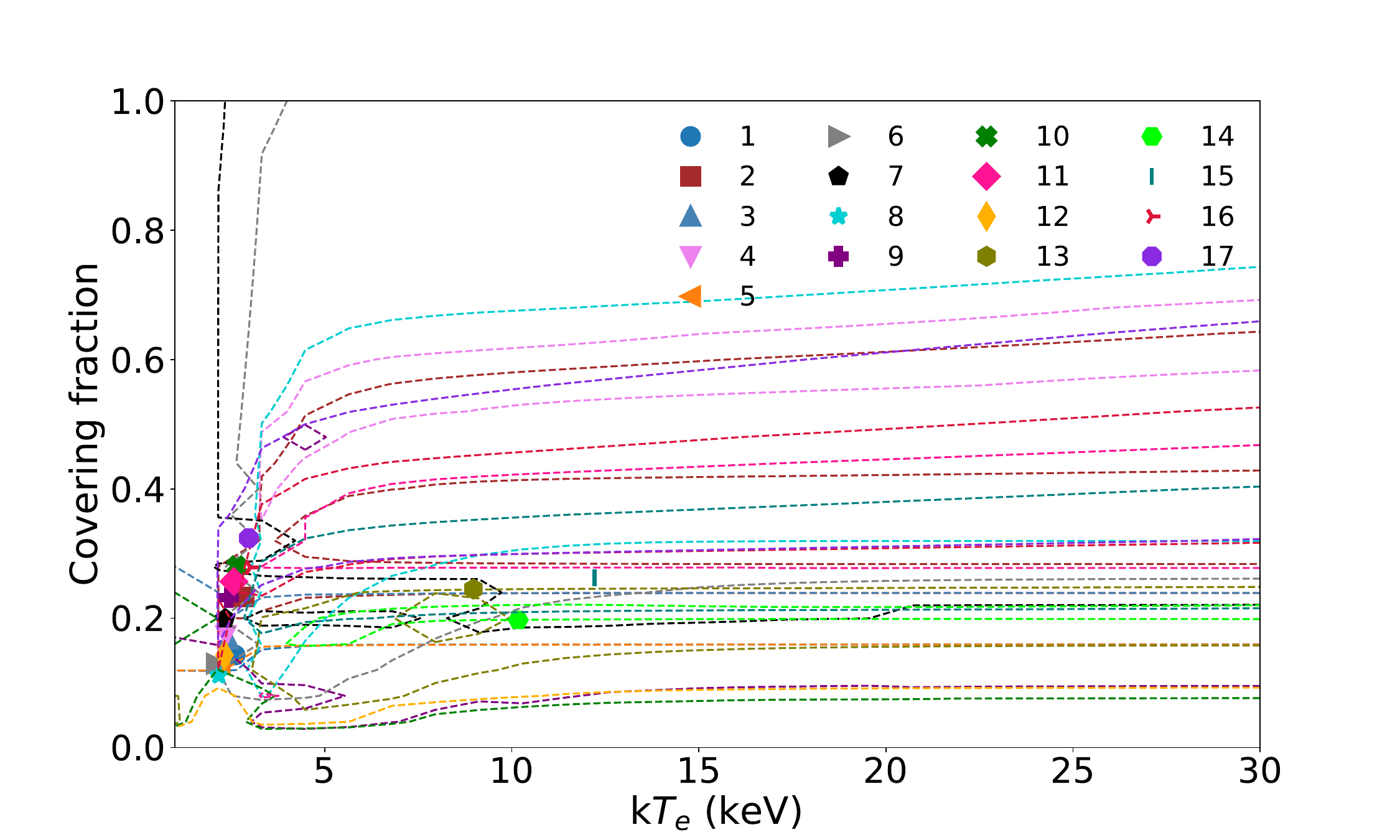}
\includegraphics[width=0.49\columnwidth]{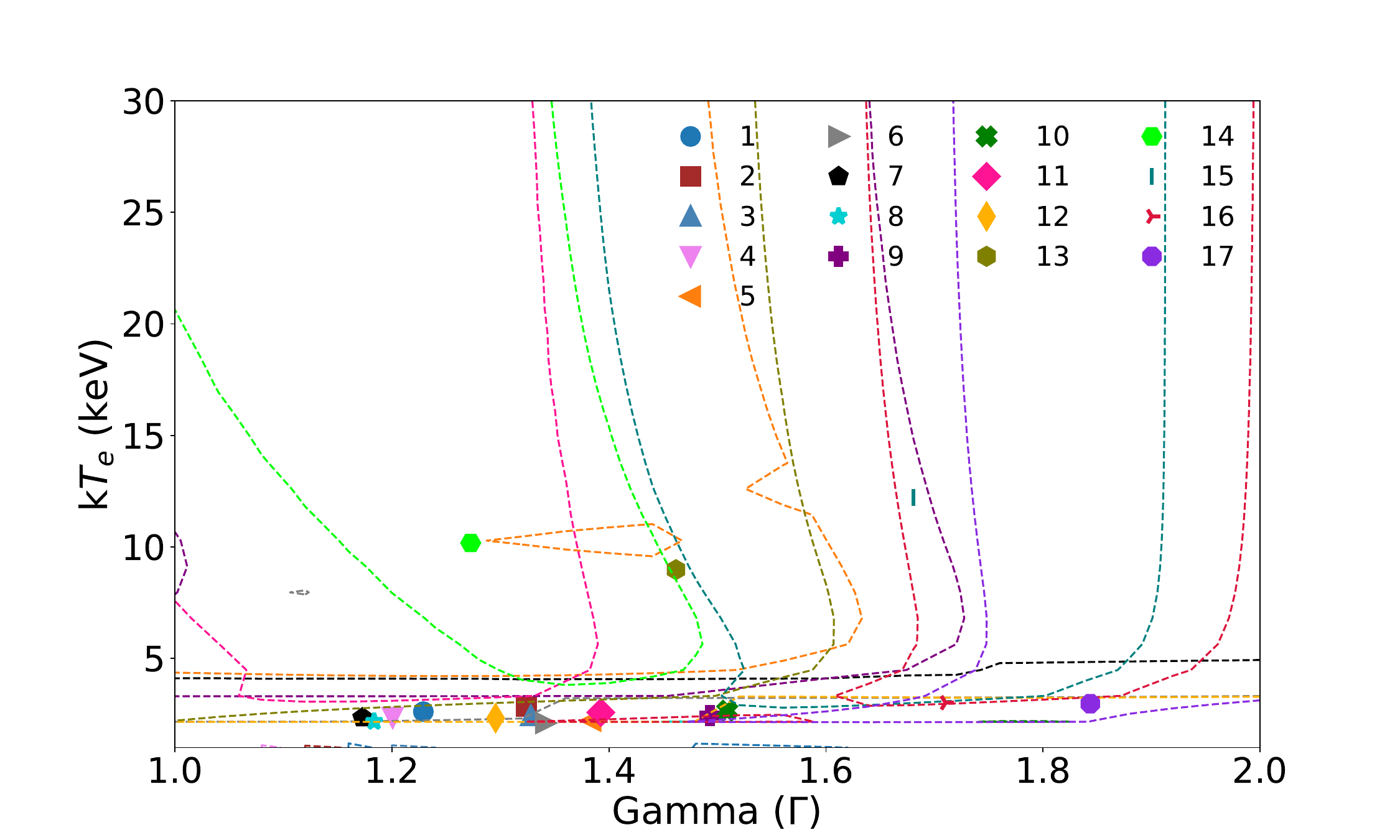}
    \caption{The above figures display the light curve and the $\chi^{2}$ contour curves for all our analyzed \textit{NICER} observations of the source SAX~J1808 during its 2022 outburst. The contour plots show the degeneracy between covering fraction (\textit{f}), $\Gamma$ and k$T_\mathrm{e}$. The contour curves correspond to the 3$\sigma$ confidence level, and each symbol shows the best-fit values for the corresponding observation. The color scheme in each contour plot is the same as in the light curve. The legend shows the Obs. no. corresponding to each observation ID, see Table~\ref{tab:NICER_TABLE}. }
    \label{fig:contours}
\end{figure*}

After deciding our primary spectral model (\texttt{tbabs*(gaussian+thcomp$\otimes$diskbb)}), we aim to study the evolution of the spectral parameters throughout the 2022 outburst of SAX~J1808. 
During the fitting process of all the \textit{NICER} spectra, we notice that a free covering fraction results in haphazard variation in the best-fit values of $\Gamma$, hinting towards a degeneracy between the two parameters. 
A similar degeneracy between these two parameters was also observed in an analysis of the \textit{NICER} observations of Aql X-1 \citep{2024MNRAS.532.3961P}, which is an NS LMXB and an intermittent AMXP. 
To get a further evidence of this degeneracy, we estimate the $\chi^2$ statistic for various combinations of $\Gamma$ and \textit{f} for {\it NICER} spectra. 
We use the \texttt{steppar} command in \texttt{XSPEC} and sample the $\Gamma$ values from 1 to 2 and \textit{f} values from 0 to 1 in a grid of linearly spaced 25 points for each parameter and estimate the $\chi^2$ statistic for each fit. 
The distribution of the $\chi^2$ statistic is shown as a contour plot in Figure~\ref{fig:contours} with dotted curves showing the $3\sigma$ confidence level.  
For many observations, one can find that various combinations of $\Gamma$ and \textit{f} lead to similar $\chi^2$ values.  

This correlation between $\Gamma$ and \textit{f} motivated us to perform the spectral analysis in two ways: by keeping the covering fraction ($f$) free, and by freezing it at various values. Figure~\ref{fig:par_var} shows the evolution of parameter values for all the considered {\it NICER} observations. 
The black line in each panel of Figure~\ref{fig:par_var} (except panel (a)) shows the evolution of individual parameters when \textit{f} is kept free during fitting. Figure~\ref{fig:par_var}(g) shows the best-fit \textit{f} value for all the observations, and Figure~\ref{fig:par_var}(a) shows the evolution of the unabsorbed flux in the energy range 0.5--10.0\,keV.

Figures~\ref{fig:par_var}(e) and \ref{fig:par_var}(g) show the haphazard variations of $\Gamma$ and \textit{f} throughout the outburst.
Considering that such variations are not physical, we investigate if a constant covering fraction can produce acceptable fits across various observations.
Thus, we take five values for $f$: 0.1, 0.15, 0.2, 0.25 and 0.3. In Table~\ref{tab:NICER_spec_pars}, we present the best-fit values of some parameters throughout the outburst using {\it NICER} data and $f=0.2$. 
This table gives a simple idea of the spectral properties and their evolution for SAX~J1808.
Above $f=0.3$, k$T_\mathrm{e}$ value becomes much larger than \textit{NICER}'s operational energy range and $\Gamma$ approaches 1 (which implies a Comptonizing medium of infinite optical depth). 
We plot the evolution of each parameter throughout the outburst for a value of \textit{f} in a distinct color (Figure~\ref{fig:par_var}). 
We fit the model \texttt{tbabs*(gaussian+thcomp$\otimes$diskbb)} in 13 of the 17 observations (from 5050260101 to 5050260106 and from 5574010101 to 5574010107), where a $\sim 1$~keV emission feature \citep{2023MNRAS.519.3811S, 2024arXiv240702360C} is fitted with a Gaussian component.  
For the rest four observations from 5574010108 to 5574010111, the model \texttt{tbabs*(thcomp$\otimes$diskbb)} is used, as there is no significant feature at $\sim 1$~keV having f-test score of $\sim6$. The 3$\sigma$ upper limit of the gaussian norm for all these four ObsIDs is $\sim$3.1$\times 10^{-4}$. We also try to fit the residual between 6--9\,keV corresponding to the iron emission complex, but find no significant improvement in the fit statistics giving $\Delta\chi^2\approx1$. The 3$\sigma$ upper limit of the gaussian norm for all the observations varies between $10^{-5}-10^{-6}$.
In some observations, the covering fraction of 0.25 or 0.3 leads to $\Gamma \approx 1$. Such cases are not depicted in Figure~\ref{fig:par_var}. 

\begin{table*}
    \centering
    \caption{The best-fit spectral parameter values for the observations of SAX~J1808 during its 2022 outburst with \textit{NICER} (section~\ref{subsec:preliminary}).}
    \label{tab:NICER_spec_pars}
    
    \begin{tabular}{|c|c|c c|c c|c c|c|c|}
    
        \hline
        \multicolumn{2}{|c|}{}& \multicolumn{2}{c|}{thcomp} & \multicolumn{2}{c|}{diskbb} & \multicolumn{2}{c|}{Gaussian} & & \\ \hline
        Obs.ID$^{\S}$ & $n_\mathrm{H}^{\ddag}$ & $\Gamma^{\P}$ & $\mathrm{k}T_\mathrm{e}^{\bullet}$\ & $\mathrm{k}T_\mathrm{in}^{\circ}$\ & norm$^{\ast}$ & LineE (keV) & norm$^{\dagger}$ & Flux$^{\#}$ & $\chi^2/$dof \\
        \hline
        \hline
        5050260101 & 7.6$\pm0.2$& 1.30$\pm0.01$ & 2.9$^{+0.2}_{-0.1}$ & 0.90$\pm0.01$ & 40$\pm2$ & 0.98$\pm0.02$ & 1.6$\pm0.2$ & 7.22$^{+0.02}_{-0.02}$ & 126/135 \\
        5050260102 & 7.7$\pm0.2$ & 1.322$\pm0.003$ & 2.7$\pm0.1$ & 0.89$\pm0.01$ & 40$\pm2$ & 0.98$\pm0.02$ & 1.6$\pm0.2$ & 7.46$^{+0.02}_{-0.02}$ & 100/143\\
        5050260103 & 7.9$\pm0.2$ & 1.343$\pm0.004$ & 2.5$\pm0.1$ & 0.88$\pm0.01$ & 44$\pm2$ & 0.98$\pm0.01$ & 1.8$\pm0.2$ & 7.71$^{+0.02}_{-0.02}$ & 93/138\\
        5050260104 & 7.8$\pm0.2$ & 1.331$\pm0.002$ & 2.3$\pm0.1$ & 0.89$\pm0.01$ & 46$\pm2$ & 0.98$\pm0.02$ & 1.5$\pm0.2$ & 7.62$^{+0.02}_{-0.02}$ & 135/143\\
        5050260105 & 6.6$\pm0.2$ & $1.34\pm0.01$ & 2.4$\pm0.1$ & 0.86$\pm0.01$ & 46$\pm2$ & 1.00$\pm0.02$ & 1.4$\pm0.2$ & 7.40$^{+0.02}_{-0.02}$ & 130/144\\
        5574010101 & 4.4$\pm0.3$& 1.34$\pm0.01$ & 2.4$\pm0.2$ & 0.86$\pm0.02$ & 48$\pm4$ & 1.03$\pm0.02$ & 1.4$\pm0.3$ & 7.53$^{+0.04}_{-0.04}$ & 127/121\\
        5574010102 & 3.1$\pm0.2$& $1.35\pm0.01$ & 2.4$\pm0.1$ & 0.87$\pm0.01$ & 46$\pm2$ & 1.04$\pm0.02$ & 1.3$\pm0.2$ & 7.54$^{+0.02}_{-0.02}$ & 147/139 \\
        5050260106 & 3.6$\pm0.3$ & 1.34$\pm0.01$ & 2.3$\pm0.1$ & 0.85$\pm0.02$ & 50$\pm3$ & 1.03$\pm0.02$ & 1.4$\pm0.3$ & 7.52$^{+0.03}_{-0.03}$ & 156/125\\
        5574010103 & 5.1$\pm0.2$ & 1.372$\pm0.004$ & 2.3$\pm0.1$ & 0.85$\pm0.01$ & 45$\pm2$ & 1.03$\pm0.02$ & 1.1$\pm0.2$ & 6.65$^{+0.01}_{-0.01}$ & 122/142 \\
        5574010104 & 9.8$\pm0.2$ & $1.42\pm0.01$ & 2.3$\pm0.1$ & 0.77$\pm0.01$ & 44$\pm2$ & 1.02$\pm0.03$ & 0.5$\pm0.2$ & 4.30$^{+0.01}_{-0.01}$ & 93/137 \\
        5574010105 & 13.6$\pm0.3$ & $1.4^{+0.2}_{-0.1}$ & 3.4$^{+0.3}_{-0.2}$ & 0.67$\pm0.01$ & 39$\pm2$ & 0.99$\pm0.05$ & 0.3$\pm0.1$ & 2.56$^{+0.05}_{-0.05}$ & 78/133\\
        5574010106 & 13.2$\pm0.2$ & 1.47$\pm0.01$ & 2.6$\pm0.1$& 0.71$\pm0.01$ & 45$\pm2$ & 0.97$\pm0.02$ & 0.5$\pm0.1$ & 3.15$^{+0.01}_{-0.01}$ & 118/135\\
        5574010107 & 15.6$\pm0.3$ & $1.5\pm0.1$ & 3.9$^{+0.5}_{-0.4}$ & 0.65$\pm0.01$ & 41$\pm2$ & 0.96$\pm0.03$ & 0.3$\pm0.1$ & 2.10$^{+0.01}_{-0.01}$ & 100/127\\
        5574010108 & 17.2$\pm0.5$ & $1.27\pm0.01$ & 11$\pm2$ & $0.56\pm0.01$ & 37$\pm3$ & - & 0.3$^{\ast\ast}$ & 9.65$^{+0.01}_{-0.01}$ & 108/126\\
        5574010109 & 15.6$\pm0.4$ & 1.59$\pm0.01$ & 3.1$^{+0.6}_{-0.4}$ & 0.57$\pm0.01$ & 37$\pm2$ & - & 0.3$^{\ast\ast}$ & 1.05$^{+0.01}_{-0.01}$ & 102/118\\
        5574010110 & 12.1$\pm0.3$ & 1.53$\pm0.01$ & 2.4$\pm0.1$ & 0.63$\pm0.01$ & 42$\pm2$ & - & 0.3$^{\ast\ast}$ & 1.76$^{+0.01}_{-0.01}$ & 135/129\\
        5574010111 & 13.2$\pm0.4$ & 1.57$\pm0.01$ & 2.0$\pm0.1$ & 0.55$\pm0.01$ & 37$\pm3$ & - & 0.3$^{\ast\ast}$ & 0.10$^{+0.01}_{-0.01}$ & 115/117\\
        \hline
    \end{tabular}

\begin{flushleft}
Notes: \\
The results in this table are for the case the covering fraction $f$=0.2 (fixed). 1$\sigma$ error bar has been reported for all the parameters.\\
$\ast \ast$ The 3$\sigma$ upper limit of gaussian norm for all four ObsIDs.\\
$\S$: The \texttt{XSPEC} model \texttt{tbabs*(gaussian+thcomp$\otimes$diskbb}) is applied to first 13 Obs.IDs (from 5050260101 - 5050260106 and 5574010101 - 5574010107) and \texttt{tbabs*(thcomp$\otimes$diskbb}) is applied to the last four observation IDs (5574010108 - 5574010111).  \\
$\ddag$: Value for the hydrogen column density ($10^{20}\mathrm{cm}^{-2}$).\\
$\P$: Photon index for \texttt{thcomp} model.\\
$\bullet$: Electron temperature for \texttt{thcomp} model (keV).\\
$\circ$: Temperature at inner disk radius (keV).\\
$\ast$: Defined as $(R_{\rm in}/D_{10})^2 \cos\theta$, where $R_{\rm in}$ is an ``apparent" inner disk radius in km, $D_{10}$ the distance to the source in units of 10\, kpc, and $\theta$ the angle of the disk ($\theta= 0$ is face-on).  \\
$\dagger$: Total photons/cm$^2$/s in the Gaussian line ($10^{-3}$).\\
$\#$: Unabsorbed flux computed in the energy range 0.5--10.0\,keV ($10^{-10}$ $\mathrm{ergs~cm^{-2}s^{-1}}$).\\
 
  \end{flushleft}
        
\end{table*}

\subsection{Spectral evolution}
\label{subsec: spectral evolution}

Here, we primarily present the {\it NICER} spectral results because our aim is to study the evolution of
spectral parameters, and the {\it NICER} observations cover the entire outburst. 
Nevertheless, in order to check the reliability of our spectral results, we compare the best-fit parameter values for the soft {\it NICER} spectrum with those for the broadband {\it NICER}+{\it AstroSat}/LAXPC (0.5--20\,keV, the background counts dominate after $\sim$20\,keV in {\it AstroSat}/LAXPC data) spectrum using the {\it NICER} ObsIDs 5574010104 and 5574010105, which are contemporaneous with the {\it AstroSat} observations (see Figure~\ref{fig:lightcurve_hid} for the overlap).
For this, we use the same model \texttt{tbabs*(gaussian+thcomp$\otimes$diskbb)}. We initially fit the \textit{NICER} spectrum with this model keeping $f$ free. 
We take this best-fit value of $f$ and freeze it while fitting the \textit{NICER}+\textit{AstroSat}/LAXPC spectrum, as keeping it free results in unphysical values of $\Gamma$ (which tends to value 1) and $\mathrm{k}T_\mathrm{e}$ (which gets pegged to the highest value of temperature range taken).
We find that the best-fit parameter values from \textit{NICER} spectral fitting and \textit{NICER}+\textit{AstroSat}/LAXPC spectral fitting are consistent with each other and differ by 1$\sigma$. For example, values for $n_\mathrm{H}$, $\Gamma$, $\mathrm{k}T_\mathrm{e}$ and $\mathrm{k}T_\mathrm{in}$ are (13.1$\pm0.4$)$\times10^{20}$\,cm$^{-2}$, 1.62$\pm0.03$, $4.2^{+1.9}_{-0.8}$\,keV and 0.68$\pm0.01$\,keV, respectively for only \textit{NICER} observation for the ObsID 5574010105 (part of the data which are contemporaneous with the \textit{AstroSat} observations).
The values for the same parameters (in same order) for \textit{NICER}+\textit{AstroSat}/LAXPC spectrum are (12.9$\pm0.3$)$\times10^{20}$\,cm$^{-2}$, 1.63$\pm0.01$, 3.8$\pm0.1$\,keV and 0.69$\pm0.01$\,keV respectively. 

\begin{figure*}
    \centering
    \subfloat{\includegraphics[width=0.49\columnwidth]{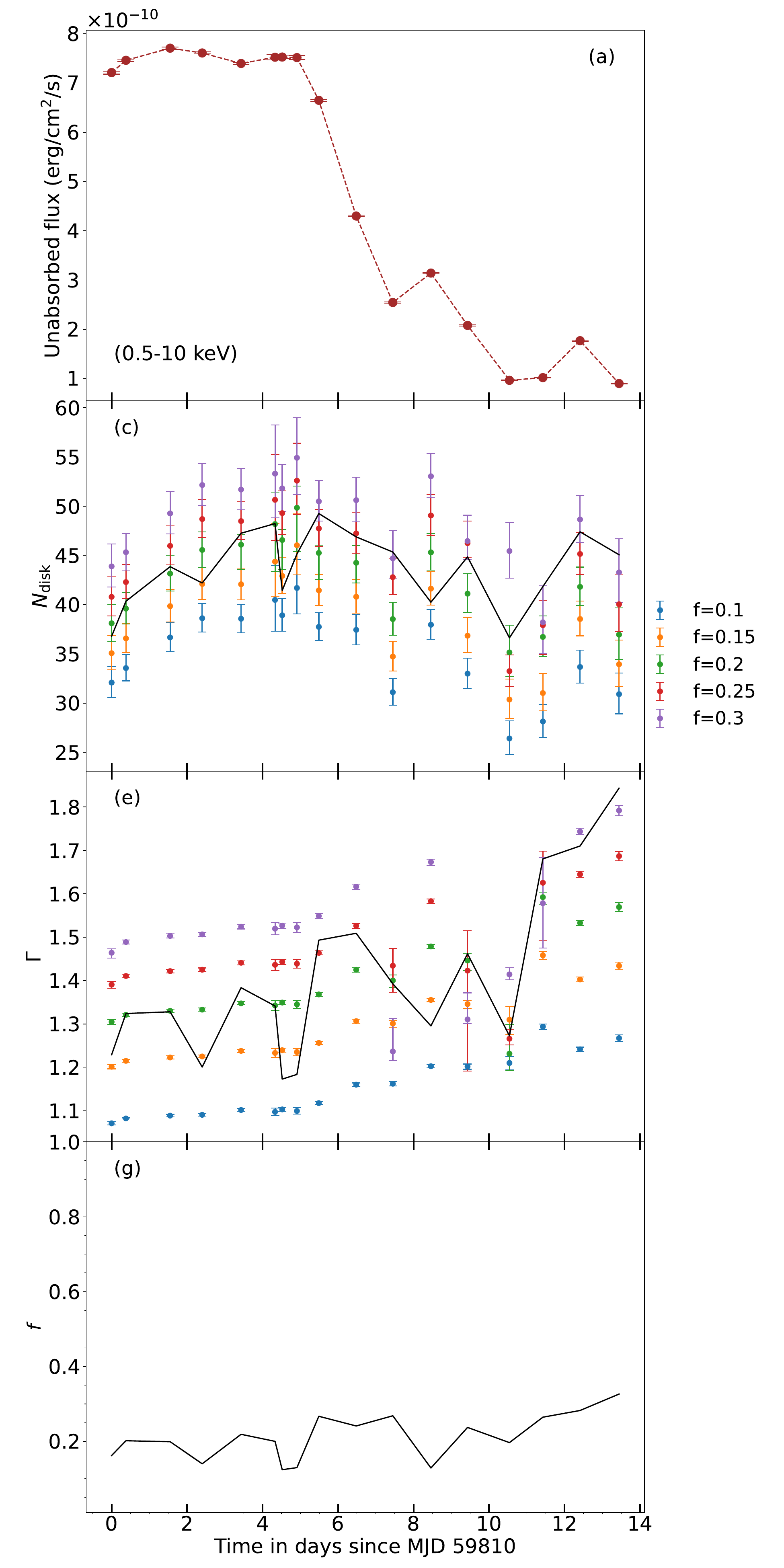}}
    \subfloat{\includegraphics[width=0.49\columnwidth]{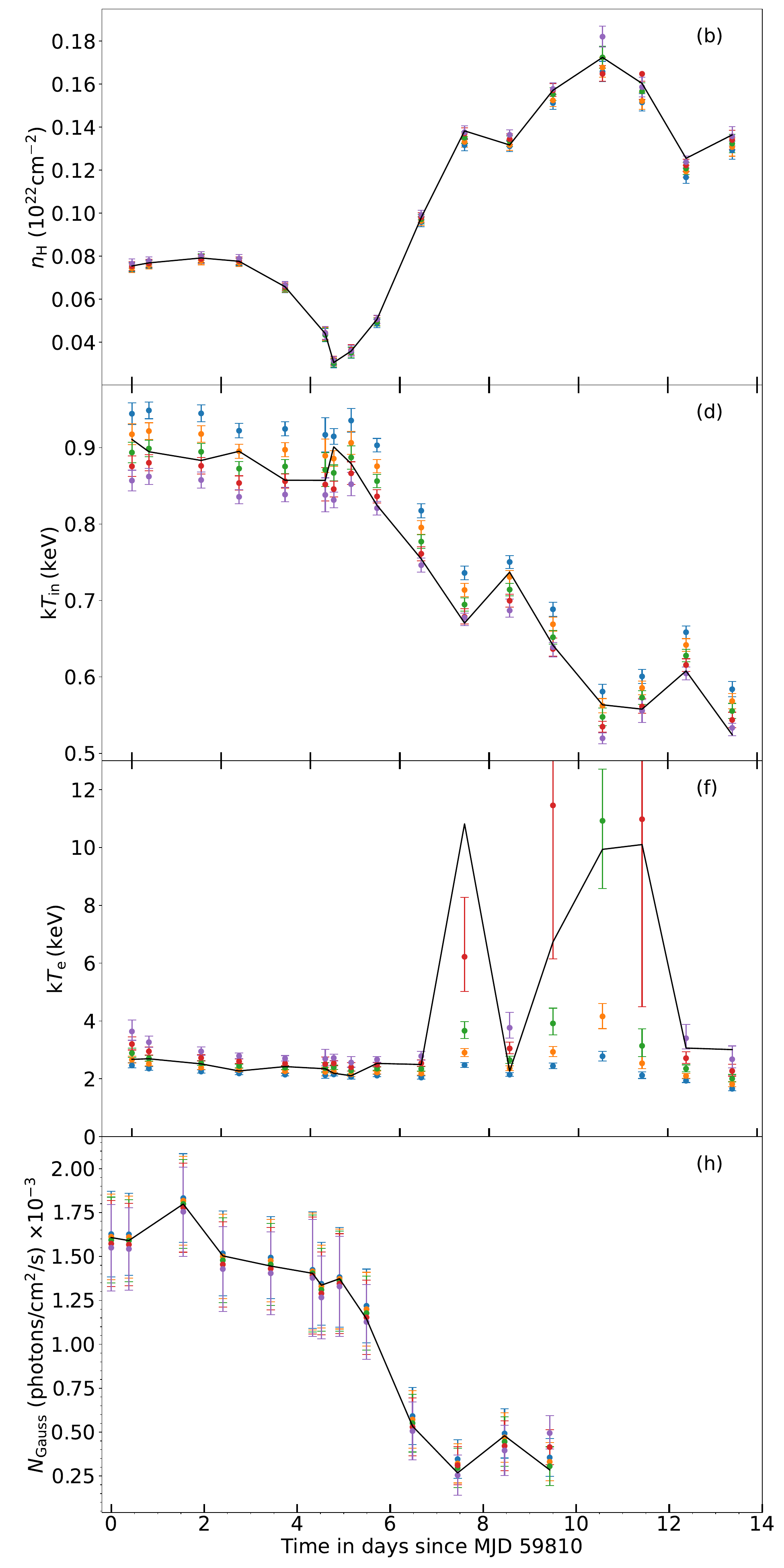}}
\caption{Evolution of best-fit spectral parameter values throughout the 2022 outburst of SAX~J1808 using the \textit{NICER} data (see section~\ref{subsec: spectral evolution} for discussion). 
The \texttt{XSPEC} model \texttt{tbabs*(gaussian+thcomp$\otimes$diskbb}) is applied to the first 13 ObsIDs (5050260101$-$5050260106 and 5574010101$-$5574010107) and \texttt{tbabs*(thcomp$\otimes$diskbb}) is applied to the last four ObsIDs (5574010108$-$5574010111). 
Each panel (except (a) and (g)) shows 
five plots using symbols in different colors, each of which is for a frozen covering fraction ($f$) value (considered values: 0.1, 0.15, 0.2, 0.25, 0.3). 
The black curve in the panels (except (a)) shows the evolution of the parameter when $f$ is kept free during the spectral fitting.
{\it Panel (a):} The evolution of the unabsorbed flux ($0.5-10$~keV) across the observations. {\it Panel (b):} The evolution of the absorption column density ($n_\mathrm{H}$). {\it Panel (c), (d):} The evolution of disk normalization ($N_\mathrm{disk}$) and  disk inner edge temperature ($\mathrm{k}T_\mathrm{in}$), respectively.
{\it Panel (e), (f):} The evolution of Comptonization photon index ($\Gamma$) and electron temperature 
(k$T_\mathrm{e}$; y-axis is limited to $\sim 12$~keV), respectively.
{\it Panel (g):} The evolution of 
the best-fit covering fraction ($f$) values. 
{\it Panel (h):}
The evolution of 
the Gaussian normalization ($N_\mathrm{Gauss}$) values. 1$\sigma$ error bar has been reported in all the above panels.} 
\label{fig:par_var}
\end{figure*}

Figure~\ref{fig:par_var}
shows that best-fit values of some parameters significantly depend on the covering fraction ($f$), e.g. $N_{\mathrm{disk}}$, $\Gamma$, while others do not, e.g. $n_{\mathrm{H}}$, $N_{\mathrm{Gauss}}$.
But, interestingly, even for a parameter which depends on $f$, the trend of its evolution along the outburst remains similar when different fixed values of $f$ are considered.
We also note that the above trend is typically and roughly followed by the black curve (corresponding to spectral fitting with free $f$) in spite of its variation.
Thus, the overall trend of the evolution of a parameter generally does not depend on $f$.
This increases the robustness of our results of spectral evolution.

The dependence on the fixed $f$ value is the strongest for 
the best-fit values of $\Gamma$ (Figure~\ref{fig:par_var}(e)). 
Yet, as mentioned above, the evolution of $\Gamma$ is very similar for every $f$ value in the considered range.
The latter is also true for $N_{\rm disk}$ (disk normalization), k$T_{\rm in}$, k$T_{\rm e}$, $n_{\rm H}$, and $N_{\rm Gauss}$ (Gaussian normalization).
We notice that k$T_{\rm in}$ evolve in a similar way the source flux does, and this trend is very clear over and above the range of the best-fit values corresponding to a range of fixed $f$ values (Figures~\ref{fig:par_var}(d)).
This trend is also clear from the black curve of the figure.
On the other hand, $n_{\rm H}$ first remains at the same level and then decreases during the top level of the source flux, and then overall increases as the flux decays (Figure~\ref{fig:par_var}(b)). 
The dependence of $n_{\rm H}$ on $f$ is also very low.

Let us now discuss the dependence of various best-fit parameter values on the frozen $f$ values from Figure~\ref{fig:par_var}. 
For a higher $f$ value, $\Gamma$ is higher (see Figure~\ref{fig:par_var}(e)), implying that the emission from corona is softer, or, less number of scattering of seed photons in corona. This is expected for a lower optical depth or a larger geometrical extension of corona, which can happen for a higher $f$.
Moreover, a less number of scattering typically means less energy is extracted from a given electron, implying that it does not cool quickly and its temperature is higher. This is consistent with a higher k$T_{\rm e}$ for a higher $f$ (see Figure~\ref{fig:par_var}(f)).
Besides, a higher $f$ implies that a larger portion of the inner disk is covered by a more extended corona, which is consistent with a higher $N_{\rm disk}$ value (see Figure~\ref{fig:par_var}(c)).
And, when the visible inner edge of the disk is farther, k$T_{\rm in}$ is expected to be smaller, as the disk blackbody temperature decreases with the radial distance. This is consistent with a lower k$T_{\rm in}$ for a higher $f$ (see Figure~\ref{fig:par_var}(d)).

\section{Aperiodic timing} \label{sec:aperiodic}

\subsection{Extraction and modeling of PDS}
We extract the root mean square (RMS)-normalized and Poisson-noise subtracted power density spectrum (PDS) using \texttt{powspec} and taking \texttt{norm=-2} (of \textsc{heasoft} v6.33) in the energy range 0.5--10\,keV for \textit{NICER} and 3--20\,keV for \textit{AstroSat}/LAXPC20. 
The light curves are binned at 2.5\,ms and segments of 328\,s are used to extract PDSs for each segment. We average and geometrically re-bin the PDSs with a factor of 1.05. 
Figure~\ref{fig:laxpc_powspec} shows the fitted power spectrum for \textit{AstroSat}/LAXPC and \textit{NICER} observations generated using contemporaneous data in the frequency range 0.004--200\,Hz. 
Neither any significant power is observed beyond this range nor we find a HFQPO (see reports for detection of HFQPO: \citealt{1998ApJ...507L..63W, 2003Natur.424...44W, 2005ApJ...619..455V, 2015ApJ...806...90B}.) Each PDS is modeled with a combination of two Lorentzian functions \citep{2000MNRAS.318..361N}.
Such a function can be written  as,

\begin{equation}
    P(\nu) = \frac{r^{2}\Delta}{2\pi} \frac{1}{(\nu - \nu_{0})^{2} + (\Delta/2)^{2}},
\end{equation}
where, $\nu_{0}$ is the centroid frequency and $\Delta$ is the full-width at half-maximum (FWHM) of the peak \citep{2002ApJ...572..392B}.
Here, $r$ is the integrated fractional RMS amplitude. 
The characteristic frequency of a Lorentzian component is $\nu_{c} =\sqrt{\nu_{0}^{2} + (\Delta/2)^{2}}$. We detect no narrow features at low frequencies that can be classified as a QPO in the spectra.
Using the same process mentioned above, we generate the PDSs for each \textit{NICER} observation in the energy ranges: 0.5--10.0\,keV, 0.5--1.5\,keV and 1.5--10.0\,keV. The best-fit Lorentzian parameters for 0.5--10.0\,keV are provided in Table~\ref{tab:Lorentzian_fit}.

\begin{table*}
    \centering
\caption{The best-fit parameter values of the Lorentzians (L: lower frequency; H: higher frequency) fitted to the PDSs ($0.5-10.0$~keV) of the {\it NICER} data from the 2022 outburst of SAX~J1808 (see section~\ref{sec:aperiodic}).}
    \label{tab:Lorentzian_fit}
    \begin{tabular}{|c|c|c|c|c|c|c|}
    
        \hline
        Obs-ID & Component & Centroid freq., $\nu_{0}$ (Hz) & FWHM, $\Delta$ (Hz) & Char. freq.$^{*}$, $\nu_{c}$ (Hz) & RMS amp. (\%) & $\chi^{2}$/dof \\
        \hline
        \hline
        5050260101& L & 0.16$\pm0.04$ & 1.4$\pm0.1$ & $0.71\pm 0.03$ & $20.0{\pm 0.2}$ & 164/160\\
                  & H & $0^{\mathrm{fixed}}$ & $63^{+14}_{-11}$ & $31\pm 6$ & $16\pm 1$ & \\
        \hline
        5050260102& L & $0^{\mathrm{fixed}}$ & 1.42$\pm0.03$ & $0.71\pm 0.01$ & $20.0\pm 0.2$ & 264/161\\
                  & H & $0^{\mathrm{fixed}}$ & 51$\pm6$ & $26\pm 3$ & $14.1\pm 0.4$ & \\
        \hline
        5050260103& L & 0.23$\pm0.02$ & 1.2$\pm0.1$ & $0.63\pm 0.02$ & $20.1\pm 0.3$ & 227/160\\
                  & H & $0^{\mathrm{fixed}}$ & 18$\pm3$ & $9\pm 2$ & $13.0\pm 0.4$ & \\
        \hline
        5050260104& L & 0.17$\pm0.04$ & 1.3$\pm0.1$ & $0.68\pm 0.03$ & $20.1\pm0.3$ & 241/160\\
                  & H & $0^{\mathrm{fixed}}$ & $32^{+13}_{-9}$ & $16\pm 5$ & $13.2\pm 0.4$ &  \\
        \hline
        5050260105& L & 0.17$\pm0.03$ & 1.7$\pm0.1$ & $0.86\pm 0.03$ & $20.2\pm 0.1$ & 206/160 \\
                  & H & $0^{\mathrm{fixed}}$ & 81$\pm10$ & $40\pm 5$ & $14.1\pm 0.4$ & \\
        \hline
        5574010102& L & 0.54$\pm0.04$ & 2.5$\pm0.1$ & $1.4\pm 0.2$ & $17.1\pm0.3$ & 209/160\\
                  & H & $0^{\mathrm{fixed}}$ & $45^{+14}_{-10}$ & $22\pm 6$ & $12.2\pm 0.4$ & \\
        \hline
        5050260106& L & 0.70$\pm0.05$ & 2.3$\pm0.2$ & $1.3\pm 0.3$ & $17.2\pm 0.3$ & 206/160\\
                  & H & $0^{\mathrm{fixed}}$ & $103^{+54}_{-38}$ & $51\pm 23$ & $14\pm 1$ & \\
        \hline
        5574010103& L & 0.38$\pm0.03$ & $2.4\pm0.1$ & $1.2\pm 0.1$ & $18.2\pm 0.1$ & 211/160\\
                  & H & $0^{\mathrm{fixed}}$ & $96^{+21}_{-17}$ & $48\pm 10$ & $13.2\pm 0.6$ & \\
        \hline
        5574010104& L & 0.24$\pm0.05$ & 2.1$\pm0.1$ & $1.1\pm 0.1$ & $18.1\pm 0.2$ & 153/160\\
                  & H & $0^{\mathrm{fixed}}$ & $92^{+19}_{-16}$ & $46\pm 9$ & $15\pm1$ & \\
        \hline
        5574010105& L & 0.17$\pm0.03$ & 1.68$\pm0.04$ & $0.86\pm 0.03$ & $20.2\pm 0.1$ & 206/160\\
                  & H & $0^{\mathrm{fixed}}$ & 81$\pm10$ & $41\pm 5$ & $14.2\pm 0.5$ & \\
        \hline
        5574010106& L & 0.20$\pm0.04$ & 1.6$\pm0.1$ & $0.84\pm 0.06$ & $17.1\pm 0.3$ & 207/160\\
                  & H & $0^{\mathrm{fixed}}$ & $58^{+13}_{-11}$ & $29\pm 6$ & $16\pm 1$ & \\
        \hline
        5574010107& L & $0^{\mathrm{fixed}}$ & 1.2$\pm0.1$ & $0.62\pm 0.03$ & $20.2\pm 0.3$ & 169/161\\
                  & H & $0^{\mathrm{fixed}}$ & $57^{+16}_{-12}$ & $29\pm 7$ & $19\pm 1$ & \\
        \hline
        5574010108& L & $0^{\mathrm{fixed}}$ & 0.93$\pm0.05$ & $0.47\pm 0.03$ & $24.1\pm 0.4$ & 141/161\\
                  & H & $0^{\mathrm{fixed}}$ & $50^{+16}_{-12}$ & $25\pm 7$ & $25\pm 2$ & \\
        \hline
        5574010109& L & $0.12^{+0.06}_{-0.08}$ & 1.3$\pm0.1$ & $0.68\pm 0.06$ & $21.2\pm 0.5$ & 192/160 \\
                  & H & $0^{\mathrm{fixed}}$ & $74^{+27}_{-19}$ & $37\pm 12$ & $24\pm 2$ & \\
        \hline
        5574010110& L & $0^{\mathrm{fixed}}$ & 2.8$\pm0.1$ & $1.4\pm 0.1$ & $18.1\pm 0.3$ & 221/163\\
        \hline
        5574010111& L & $0.2^{+0.1}_{-0.2}$ & 1.60$\pm0.03$ & $0.8\pm 0.2$ & $19\pm 1$ & 184/160\\
                  & H & $0^{\mathrm{fixed}}$ & $48^{+47}_{-18}$ & $24\pm 16$ & $26\pm 4$ &  \\
        \hline
    \end{tabular}
    \begin{flushleft}
    Notes: \\
    1$\sigma$ error bar has been reported for all the parameters.\\
    $\ast$: Characteristic frequency of a Lorentzian component is defined as $\nu_{c}=\sqrt{\nu_{0}^{2} + (\Delta/2)^{2}}$, where $\nu_{0}$ is the centroid frequency and $\Delta$ is the full-width at half-maximum (FWHM) of the peak.  \\
  \end{flushleft}
        
\end{table*}

\subsection{Results}\label{aperiodic_results}

\begin{figure*}
\centering
\subfloat(a){\includegraphics[width=0.47\columnwidth]{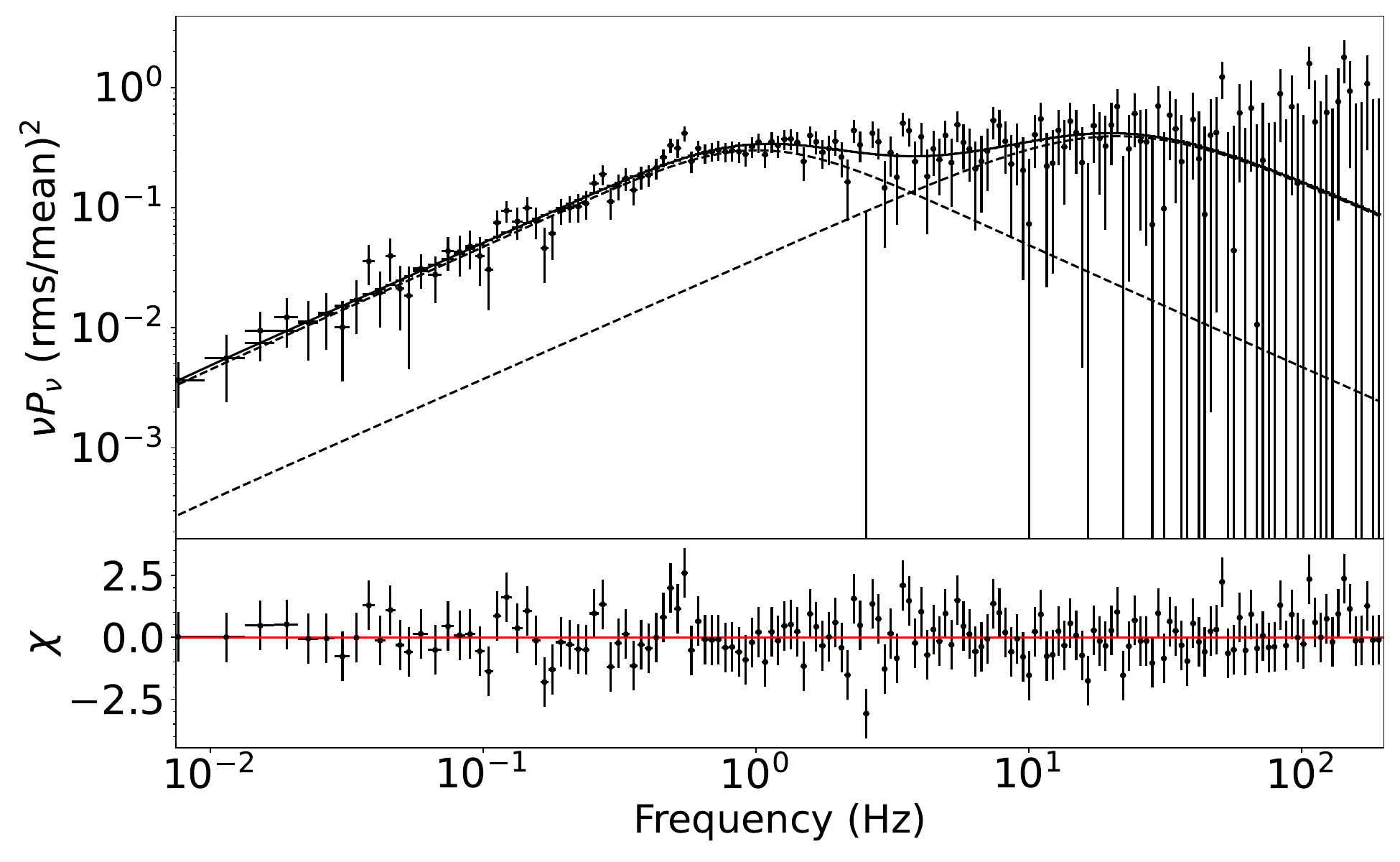}}
\subfloat(b){\includegraphics[width=0.47\columnwidth]{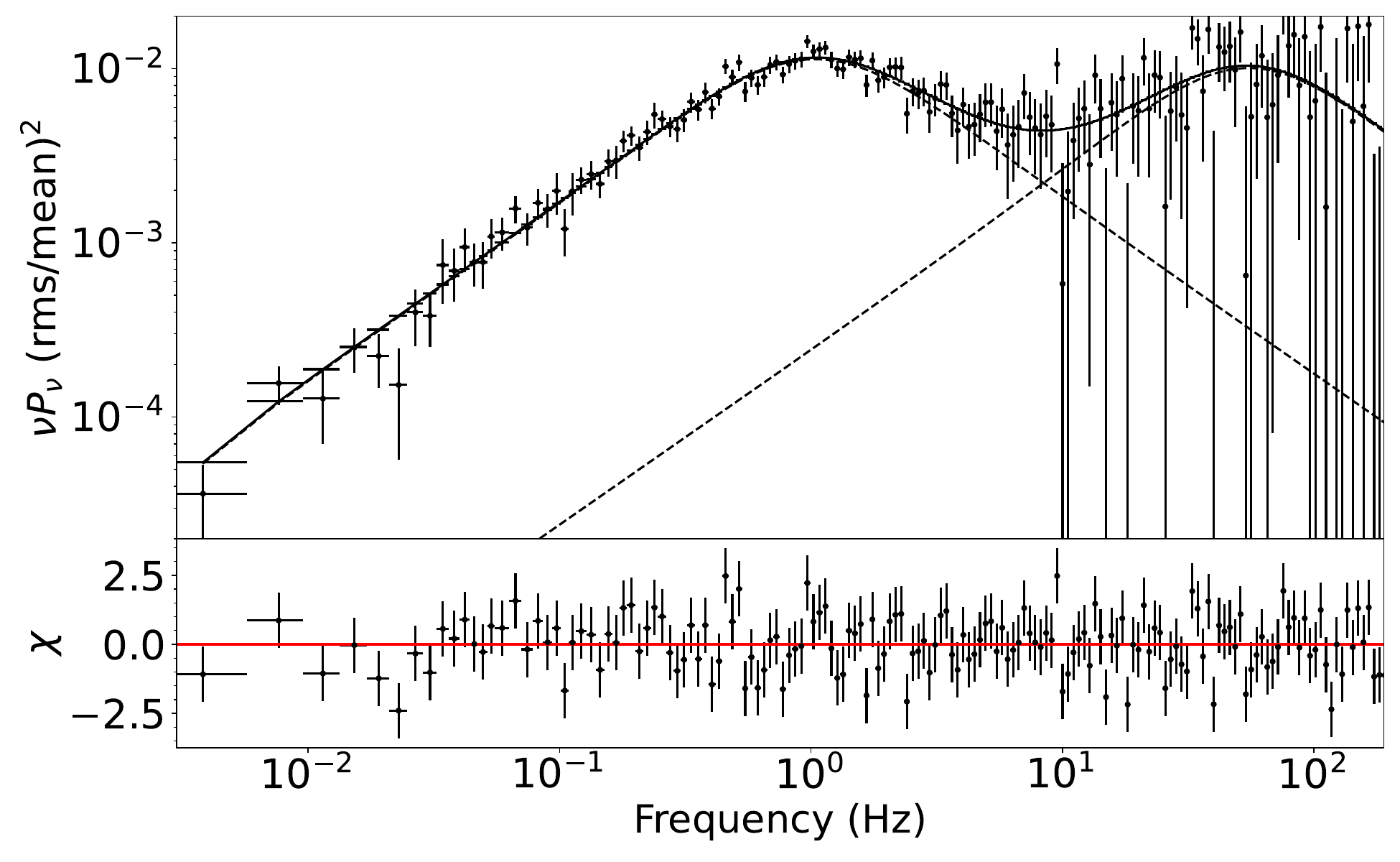}}
\caption{RMS-normalized and Poisson-noise subtracted power density spectra (PDSs) of SAX~J1808 with 1$\sigma$ errors (section~\ref{sec:aperiodic}).
Two Lorentzians are used to fit the PDSs in $0.004-200$~Hz frequency range.  
{\it Panel (a)} is for \textit{AstroSat}/LAXPC data in 3--20 keV, and {\it panel (b)} is for {\it NICER} data in 0.5--10 keV.}
\label{fig:laxpc_powspec}
\end{figure*}

In the aperiodic timing analysis of the contemporaneous \textit{NICER} and \textit{AstroSat}/LAXPC data (Figure~\ref{fig:laxpc_powspec}), we note the presence of two broad frequency components, which we denote a low or a high frequency component based on its characteristic frequency. We fit the low-frequency Lorentzian to the \textit{NICER} data having a centroid frequency of 0.3$\pm0.1$\,Hz with FWHM and characteristic frequency of 2.1$\pm0.1$\,Hz and 1.05$\pm0.01$\,Hz,  respectively. We fix the centroid frequency to zero for the high-frequency Lorentzian, \citep[e.g.][]{2021MNRAS.508.3104B} giving FWHM and characteristic frequency of $123^{+22}_{-19}$\,Hz and 62$\pm10$\,Hz, respectively. The addition of this high frequency Lorentzian gives the F-test score of $\sim68$ with probability of improvement by chance $\sim10^{-22}$.
For \textit{AstroSat}/LAXPC, both the Lorentzian components are fixed at zero centroid frequency \citep[e.g.,][]{2019MNRAS.488..720B}, and the FWHM and characteristic frequency for the low-frequency Lorentzian are observed to be 2.4$\pm0.2$\,Hz and 1.2$\pm0.8$\,Hz, respectively, and those for the high-frequency Lorentzian are estimated at $55^{+20}_{-14}$\,Hz and 27$\pm9$\,Hz,  respectively. The addition of the high frequency Lorentzian in this case gives a F-test score of $\sim66$ with probability of improvement by chance $\sim10^{-19}$.
Since the operational energy ranges for \textit{NICER} and \textit{AstroSat}/LAXPC are quite different,  we do not expect the parameters of the fitted Lorentzians in these cases to have similar values.

\begin{figure}
\centering
\includegraphics[width=0.5\columnwidth]{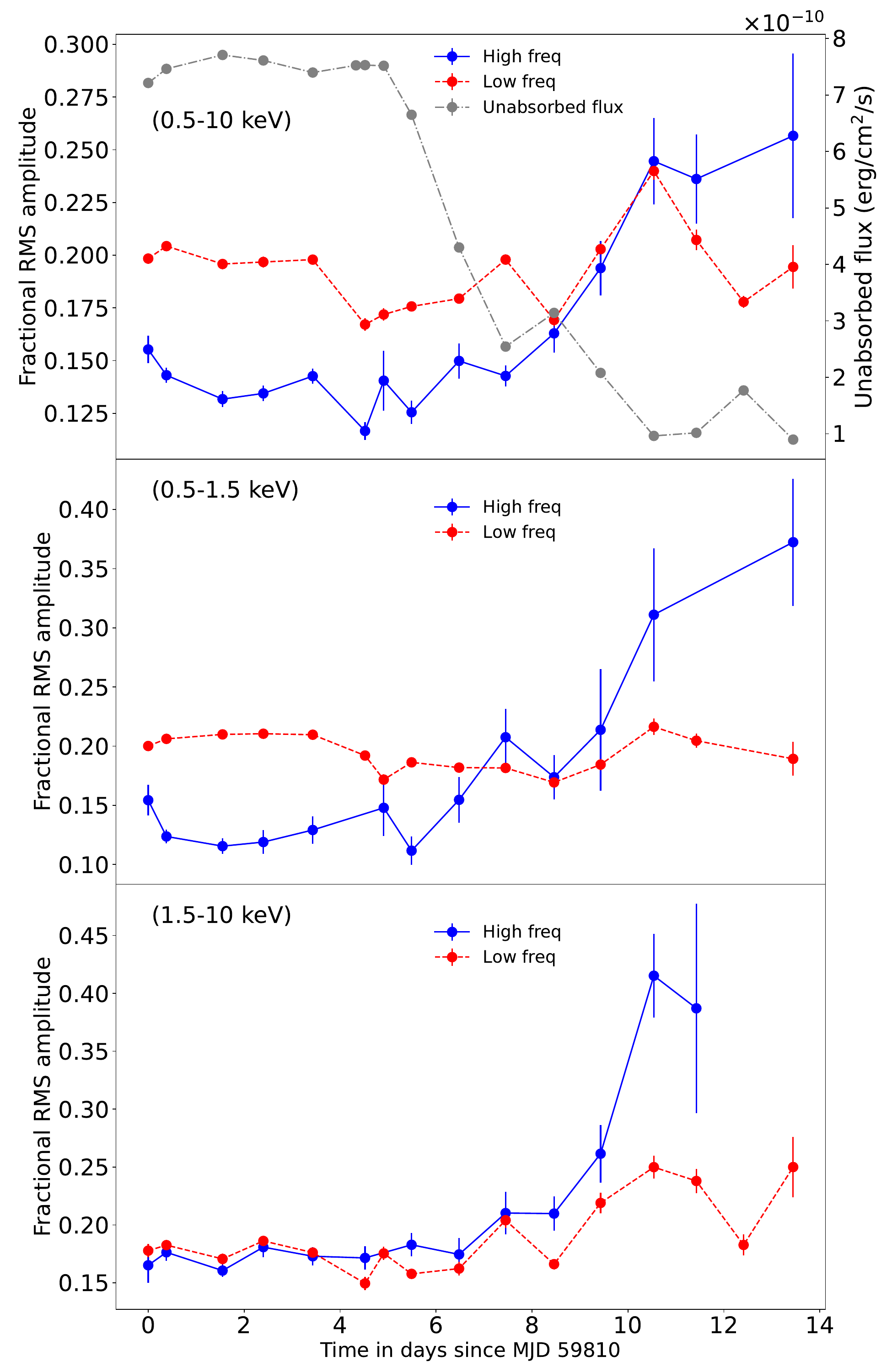}
\caption{The evolution of flux (gray, dash-dot) and fractional RMS amplitude of the
low-frequency (red, dashed) and high-frequency (blue, solid) Lorentzians
fitted to the PDSs of the \textit{NICER} data from the source SAX~J1808 during its 2022 outburst.
The three panels are for the energy ranges 0.5--10.0\,keV, 0.5--1.5\,keV and 1.5--10.0\,keV.  
1$\sigma$ error for each point is given.}
\label{fig:rms}
\end{figure}

In order to investigate the evolution of energy-dependent fractional RMS amplitude, we model the PDS of each \textit{NICER} observation with Lorentzian(s) for three energy bands: 0.5--10.0\,keV, 0.5--1.5\,keV and 1.5--10.0\,keV. 
We see that majority of the PDSs require two Lorentzians to give an acceptable fit in all the three bands.
Table~\ref{tab:Lorentzian_fit} gives the parameter values of the fitted Lorentzians in the energy range 0.5--10.0\,keV.  
Figure~\ref{fig:rms} shows the plots for the evolution of fractional RMS amplitude throughout the outburst for both low and high-frequency Lorentzian in all the three energy ranges.
We find that, during the first part of the outburst when the flux is high, low-frequency fractional RMS amplitude is significantly higher than the high-frequency fractional RMS amplitude for $0.5-10$~keV and $0.5-1.5$~keV energy ranges, while the two amplitudes are similar for $1.5-10$~keV.
When the flux declines, low-frequency amplitude remains roughly at similar levels (although with minor fluctuations) but high-frequency amplitude rises for all three energy ranges.

To get a better understanding of the emission components involved in these low and high-frequency fluctuations, we investigate the time lag caused between the hard and soft photons.
To do this we use the module \texttt{AveragedCrossspectrum} from the software \texttt{STINGRAY v2.1}. We divide the \textit{NICER} observation event files in the energy ranges 0.5--1.5\,keV (soft photons) and 1.5--10.0\,keV (hard photons). 
For our case, we take the time bin of 0.0025\,s to get a maximum frequency  of 200\,Hz as done while generating the PDS. 
We calculate the time lag between soft and hard photons in the two frequency ranges: 0.004--2\,Hz corresponding to the low-frequency Lorentzian, and 10--100\,Hz corresponding to the high-frequency Lorentzian (see Figure~\ref{fig:laxpc_powspec}(b)). 
Event files of observations are combined together using the \textit{NICER} module \texttt{niobsmerge} and divide in the energy ranges as mentioned above. 
We rebin to have a single bin for each of 0.004--2\,Hz and 10--100\,Hz.
This procedure is done for two sets of observations: the first seven observations (ObsID: 5050260101--5050260106 and 5574010102. We have not included ObsID: 5574010101 because of its low exposure time) and all the observations. Note that we need to combine frequencies and observations to improve the statistics.

In both the frequency bands and for both sets of observations, we see a hard lag, i.e., variability in the hard photons are detected later than that in the soft photons.
However, we find significantly and substantially more hard lag in the 0.004--2\,Hz frequency band than in the 10--100\,Hz band, for first 7 observations: $11.6\pm0.6$\,ms and for all observations: $10.6\pm0.5$\,ms. Whereas, the detection of hard lag for the 10--100\,Hz band is marginal, for first 7 observations: $0.21\pm0.15$\,ms and for all observations: $0.19\pm0.15$\,ms.

\section{Discussion} \label{sec:discussion}
In this paper, we investigate the evolution of spectral and aperiodic timing properties of the AMXP SAX J1808.4$-$3658 during its 2022 outburst.
We primarily use the \textit{NICER} data throughout the outburst, but also analyze the contemporaneous  \textit{AstroSat} observations.
We find that the source continuum spectra throughout the outburst are well modeled with the emissions from two components: an accretion disk and a Comptonizing corona with its seed photons supplied by the disk.
We particularly show that a disk emission explains the softer thermal emission much better than a single temperature blackbody (section~\ref{sec:spectral}).
This conclusion is crucial for further understanding of emission components and accretion processes.
The success of the same model throughout the outburst strongly suggests that the same general accretion process and physics are applicable to the entire outburst, enables us to study the outburst in a uniform  manner, and allows us to probe which parameters change as the accretion rate evolves.

The most remarkable evolution is of the absorption column density $n_{\rm H}$ (Figure~\ref{fig:par_var}(b)).
The value of $n_\mathrm{H}$ varies significantly throughout the outburst in the range $(3.1-17.2)\times10^{20}$~cm$^{-2}$ (see Table~\ref{tab:NICER_spec_pars}) consistent with previously observed values for this source \citep[e.g.,][]{2007ApJ...660.1424H, 2009MNRAS.396L..51P, 2019MNRAS.483..767D, 2023MNRAS.519.3811S}.
The very significant, substantial, and systematic variation of $n_{\rm H}$ (section~\ref{subsec:preliminary}) clearly shows that there is an evolving atomic hydrogen absorber intrinsic to SAX~J1808 in addition to the Galactic column density of atomic hydrogen because the latter should not evolve.
This intrinsic component is distinct from the Comptonizing corona, as there is almost no degeneracy between $n_{\rm H}$ evolution and  the covering fraction ($f$) of the corona (see Figure~\ref{fig:par_var}(b)).
Thus, we hypothesize that there are two absorbing/scattering components in SAX~J1808, an atomic hydrogen medium and a corona
\citep[the \textit{NICER} data from Aql X-1 also showed a similar aspect:][]{2024MNRAS.532.3961P}. 
As noted in section~\ref{subsec: spectral evolution}, there appears to be a connection between the long-term source flux evolution and the $n_{\rm H}$ evolution  during the outburst (see Figure~\ref{fig:par_var}(b)).
During first two days when the source flux was at the top,  $n_{\rm H}$ did not change much. But, during the next two days, while the source flux was still at the top, $n_{\rm H}$ reduced by a factor of 2. It is not clear why this happened, but this could be because of the long-term high X-ray flux from the source. Then, as the X-ray flux decayed in the next few days, the best-fit value of $n_{\rm H}$ rose by a factor of $5-6$.

As the accretion rate, and hence the source flux, decays from a higher to a lower value, we expect for a geometrically thin Keplerian disk that the blackbody temperature at any radius also decreases \citep{2002apa..book.....F}.
This is what we find irrespective of the frozen $f$ value from panels (a) and (d) of Figure~\ref{fig:par_var}.
The best-fit $\mathrm{k}T_\mathrm{in}$ values ($0.55–0.90$ keV; Table~\ref{tab:NICER_spec_pars}) obtained are comparable to the values observed earlier \citep{1998ApJ...506L..35H, 2009MNRAS.396L..51P, 2023MNRAS.519.3811S}.
The disk normalization ($N_{\rm disk}$) does not have a clear trend although it appears that $N_{\rm disk}$ somewhat increases during the initial part of the outburst and then overall slightly decreases (with minor variations; Figure~\ref{fig:par_var}(c); Table~\ref{tab:NICER_spec_pars}).
This suggests that the estimated disk inner edge radius ($\propto \sqrt{N_{\rm disk}}$) does not correspond to the NS magnetospheric radius and hence  should have a higher value \citep{1979ApJ...234..296G, 1991PhR...203....1B}.
Moreover, Figure~\ref{fig:par_var}(c) shows substantial degeneracy between the $N_{\rm disk}$ and the $f$ parameter.
The above two observations suggest that the inner part of the disk is substantially covered by the corona, determining the estimated disk inner edge radius, and hence the corona is somewhat compact around the inner portion of the accretion flow, and the extent of the corona does not substantially evolve throughout the outburst.
The electron temperature (k$T_{\rm e}$) of this corona does not show a clear trend with some fluctuations, (Figure~\ref{fig:par_var}(f); Table~\ref{tab:NICER_spec_pars}).
But, the Comptonization photon index ($\Gamma$) somewhat increases as the outburst decays (Figure~\ref{fig:par_var}(e)).
This could be because, as the accretion rate decreases, the density, and hence the optical depth, of the corona may also decrease. Thus, the number of scattering of seed photons may decrease, resulting in a softer coronal emission indicated by larger $\Gamma$ values.
We further note that the ranges of $\Gamma$ ($1.27-1.59$) and k$T_{\rm e}$ ($2.3-10.9$~keV) in Table~\ref{tab:NICER_spec_pars} are comparable to previously observed values \citep[e.g., ][]{ 1998ApJ...506L..35H, heise1998discovery, 2002ApJ...575L..15C, 2019MNRAS.483..767D}.

As indicated above, the estimated disk inner edge radius can provide an upper limit to the neutron star radius ($R$), radius ($R_{\rm ISCO}$) of the innermost stable circular orbit (ISCO), and the magnetospheric radius \citep[$R_{\rm mag}$; ][]{1979ApJ...234..296G, BhattacharyyaChakrabarty2017}.
Considering the maximum $N_{\rm disk}$, which is defined as $(R_{in}/D_{10})^2$ \text{cos}$\theta$, where $R_{in}$ is “an apparent” inner disk radius in km, $D_{10}$ the distance to the source in units of 10\,kpc, and $\theta$ the angle of the disk ($\theta$ = 0 is face-on), for a conservative estimation, and the source distance in the range of $3.4-3.6$~kpc \citep{2006ApJ...652..559G}, the inclination angle of 69$^\circ$ \citep{2019MNRAS.490.2228G}, and a  color factor of 1.8 
\citep{2005MNRAS.359.1261G}, we get a disk inner edge radius of $\sim14$\,km.
This, along with the Schwarzschild expression of $R_{\rm ISCO}$ ($=6GM/c^2$), gives an upper limit of the NS mass $M$ as $1.6~M_\odot$.
An upper limit of $R_{\rm mag}$ gives a lower limit of the NS magnetic field $B$ \citep[see equation 2;][]{2018MNRAS.480..692P}. 
While this lower limit depends on a number of unknown parameters such as NS mass and radius, it generally comes out to be within a previously estimated range of $B$ for SAX~J1808
\citep[$(0.14-1.77)\times10^8$~G;][]{Mukherjee1etal2015}.

\begin{equation}
    B = 1.32 \times 10^8 (G) \left( \frac{\dot{M}}{\dot{M}_{\mathrm{Edd}}} \right)^{1/2} m^{1/4} R_6^{-5/4} \phi^{-7/4}
\end{equation}

It is important to probe how the fluxes of the two continuum emission components, a disk and a corona, in various energy bands evolve with time in order to gain an insight of the accretion processes.
For this, we consider three energy bands: $0.5-10.0$~keV, $0.5-1.5$~keV and $1.5-10.0$~keV, see figure~\ref{fig:appendix_plots}.
We find that both disk and corona fluxes do not change much during the top level of the outburst, and decrease as the outburst decays.
This implies that the accretion rate determines the properties and emissions of both the disk and the corona.
In the $0.5-10.0$~keV, $0.5-1.5$~keV and $1.5-10.0$~keV energy bands,
the disk-flux/corona-flux ratio (figure~\ref{fig:appendix_plots}(d)) is about 
$\sim 1$, $\sim 5$ and $\sim 0.7$ at the top level of the outburst, and $\sim 1$, $\sim 4$ and $\sim 0.5$ towards the outburst end.
Since the corona upscatters the disk photons, it is expected that the contribution of the corona to the total flux is larger for higher energies, as we find.
The contribution of the corona to the total flux also overall increases as the outburst decays, because the disk emission decreases faster than the coronal emission in the later part of the outburst.

Let us now consider our aperiodic timing results to further understand the accretion and emission processes.
We find two broadband features, of low and high-frequencies, which are distinguished by their separate peaks (Figure~\ref{fig:laxpc_powspec}) and distinct fractional RMS amplitude behavior (Figure~\ref{fig:rms}).
The first and the most important question is which of the disk and the corona fluctuates to cause them. 
Here, we assume that the physics of origin of each of the low-frequency and high-frequency features is the same in all three energy bands, and only the parameter values can be different. This should be reasonable, because otherwise no general model of the feature will be possible.
Similarly, we assume that the physics of origin of each of these aperiodic timing features is the same throughout the outburst, and only the parameter values can be different.
This could be expected if the general accretion process and physics do not significantly evolve during the outburst.

On comparing the relative contributions of the disk and the coronal flux, we find that neither can separately account for the amplitudes of either of the features in all three energy bands throughout the entire outburst, see figure~\ref{fig:appendix_plots}.
From the plots we see, for example in the energy range 0.5-1.5 keV the Compton flux is around 10$\%$ of the total flux and hence cannot account for $>$20$\%$ variability seen in both high and low frequency, even if it varies 100$\%$. A similar argument can be made in the other energy ranges by comparing the variability and the relative flux of the two components.
For some other cases, a large variation of the coronal emission (e.g., $\sim 60-80$\%) is required to explain the observed fluctuation amplitudes. 
Thinking along the same line for disk emission, for energy band 1.5-10.0 keV the disk has to be 60--80$\%$ variable in the later part of the outburst to account for the observed RMS variability (figure~\ref{fig:appendix_plots}(b)). But, we don’t expect such large variations in flux from these extended sources. Considering intrinsic fluctuation of emission from disk or corona, a large fluctuation would mean a large structural change of the disk or corona across a large length scale. We do not know any mechanism or reason for such large structural change.
Therefore, using the argument of the previous paragraph, we conclude that both the disk and the corona contribute to each of the aperiodic timing features.

If an aperiodic broadband feature originates from both the disk and the corona, then there must be a connected mechanism operating in these two spectral components. 
Otherwise, it would not be the same timing feature. 
A natural way to explain this is that the fluctuation originates in the disk and some of these disk photons are up-scattered by the corona to cause the fluctuation of the coronal emission.
This is nicely consistent with our spectral fitting formalism, where the disk photons are up-scattered by the corona (section~\ref{subsec:preliminary}).
As indicated in section~\ref{sec:intro}, disk fluctuations  could be due to the variability of disk parameters at different radii with various time scales, which may originate at larger radii with lower frequencies and propagate inwards causing the high-frequency variability in the inner parts of the disk.
Thus, our low-frequency feature may originate at higher disk radii, while the high-frequency feature may be associated with the inner parts of the disk.
A fraction of each set of photons, from outer and inner parts of the disk, could be up-scattered by the corona. 
If this corona is somewhat compact around the inner portion of the disk, as inferred from the spectral analysis in this section, then that could naturally explain the observed larger hard lag of the low-frequency feature relative to the high-frequency feature.
This is because the hard lag may partly originate due to the path difference between the two sets of photons: those reaching the observer directly from the disk and those originated from the disk but reaching the observer after being up-scattered in the corona. 
The up-scattered photons originated from the larger disk radii with lower frequencies should take more time to reach the observer than those originated from the inner parts of the disk.
Finally, we also note that the scattering processes in the corona should also contribute to the hard lag. 
In particular, the low electron temperature and moderate photon index would imply a high optical depth of the corona, which in our case is $\sim24$ \citep[as computed using the typical values of photon index and the electron temperature and using the formula from][]{2019MNRAS.488..720B}, and is also seen in other NS LMXBs \citep[e.g.][]{2023ApJ...955..102B}. 
Multiple scatterings of the seed photons could contribute partially towards the relatively large hard lag observed here.

The above scenario is supported by Figure~\ref{fig:rms}. 
This figure shows that the low-frequency fractional RMS amplitudes are higher for lower energies during the top of the outburst, and overall does not change much throughout the outburst for lower energies.
Since the disk's relative contribution is higher during the top of the outburst and for lower energies,
the above suggests that the low-frequency feature is primarily associated with the disk. 
This is possible if this feature originates at larger disk radii so that a relatively small fraction of the corresponding photons is intercepted by a compact, centrally located corona.
On the other hand, the high-frequency fractional RMS amplitudes are higher for higher energies during the top of the outburst, and clearly increase towards the end of the outburst (Figure~\ref{fig:rms}).
Considering that the relative contribution of corona is lower during the initial part of the outburst and increases both, towards the end of the outburst and at higher energies, it suggests that the high-frequency feature is primarily associated with the corona.
This is possible if this feature originates at the inner parts of disk so that a relatively large fraction of the corresponding photons is intercepted by a compact, centrally located corona.

\section{Summary}\label{sec:summary}

We summarize the main points of this paper below.
\begin{itemize}[noitemsep, topsep=0pt]

\item We find that the {\it NICER} continuum spectra ($0.5-10.0$~keV) of the 2022 outburst of the AMXP SAX J1808.4$-$3658, can be well described with emissions from an accretion disk and a centrally located compact corona throughout the outburst.\\

\item While both the disk and the coronal emission decreased, and both these components softened, during the outburst decay, the relative contribution of the corona to the source emission was higher for higher energies and increased with the decline of the outburst.\\

\item We find an intrinsic atomic hydrogen medium in the system, which significantly, substantially, and systematically evolved throughout the outburst.\\

\item From the timing analysis, we find two broadband aperiodic features: one of low-frequency ($\sim 0.004-2$~Hz), another of high-frequency ($\sim 10-100$~Hz).
The former has $\sim 11$ ms of significant hard lag between $1.5-10.0$~keV and $0.5–1.5$~keV photons.\\

\item Considering spectral and timing results, we conclude that both the disk photons and the photons up-scattered by the corona contributed to each of the aperiodic timing features.
However, the low-frequency feature, originated at larger disk radii. was primarily associated with the disk, while the centrally located compact corona mainly contributed to the high-frequency feature. 
 
\end{itemize}

\begin{acknowledgements}
We thank the anonymous referee for thoughtful comments which significantly improved this paper. We are thankful to the \textit{NICER} team for efficient and continuous monitoring of the source. This research has made use of \textit{NICER}'s data obtained through the High Energy Astrophysics Science Archive Research Center (HEASARC) Online Service, provided by the NASA/Goddard Space Flight Center. This research has also made use of the \textit{AstroSat} data, obtained from the Indian Space Science Data Centre (ISSDC). We thank the \textit{AstroSat}/LAXPC Payload Operation Center (POC) at TIFR, Mumbai, for providing the necessary software tools.
SB acknowledges the International Space Science Institute (ISSI), Bern, Switzerland, for supporting an academic visit, which was useful for this research.
AK, YB, and SB dedicate this paper to the memory of their coauthor, the late Professor M.  Falanga, who passed away during the course of this work.
\end{acknowledgements}

\appendix 
\renewcommand{\thesection}{\Alph{section}}
\setcounter{figure}{0}
\renewcommand{\thefigure}{A\arabic{figure}}

\begin{figure*}
    \centering
\includegraphics[width=0.45\columnwidth]{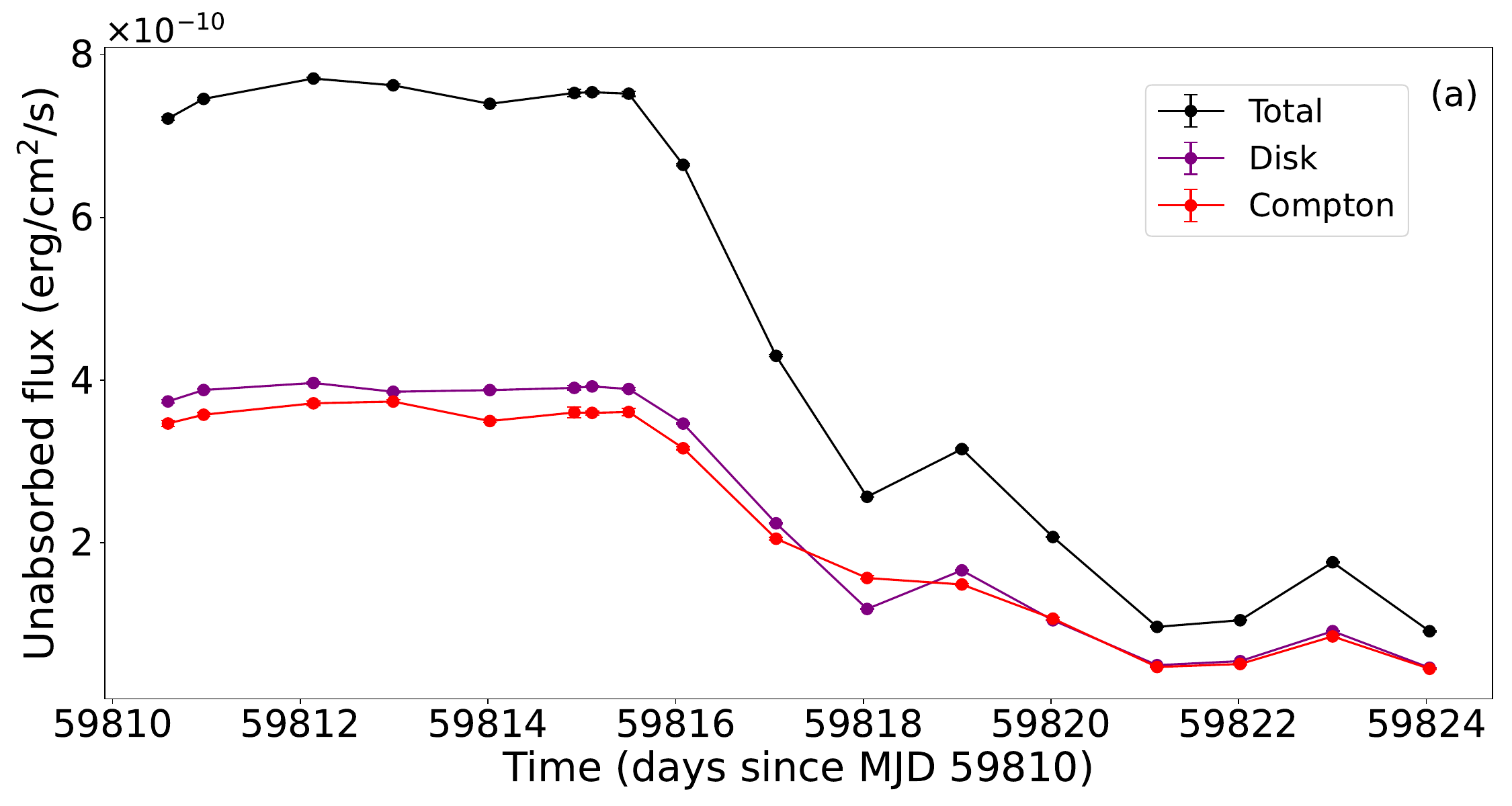}
\includegraphics[width=0.45\columnwidth]{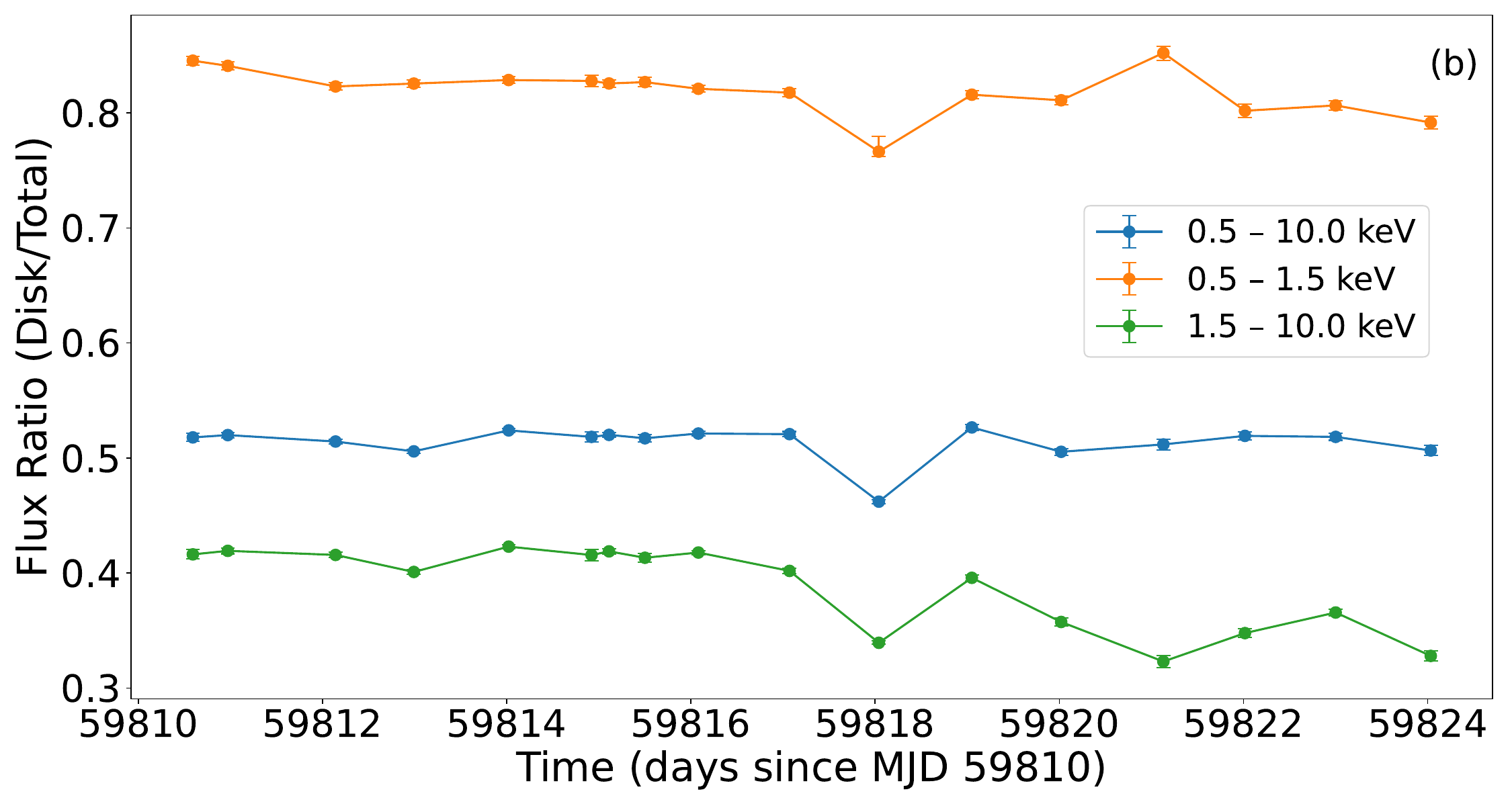}
\includegraphics[width=0.45\columnwidth]{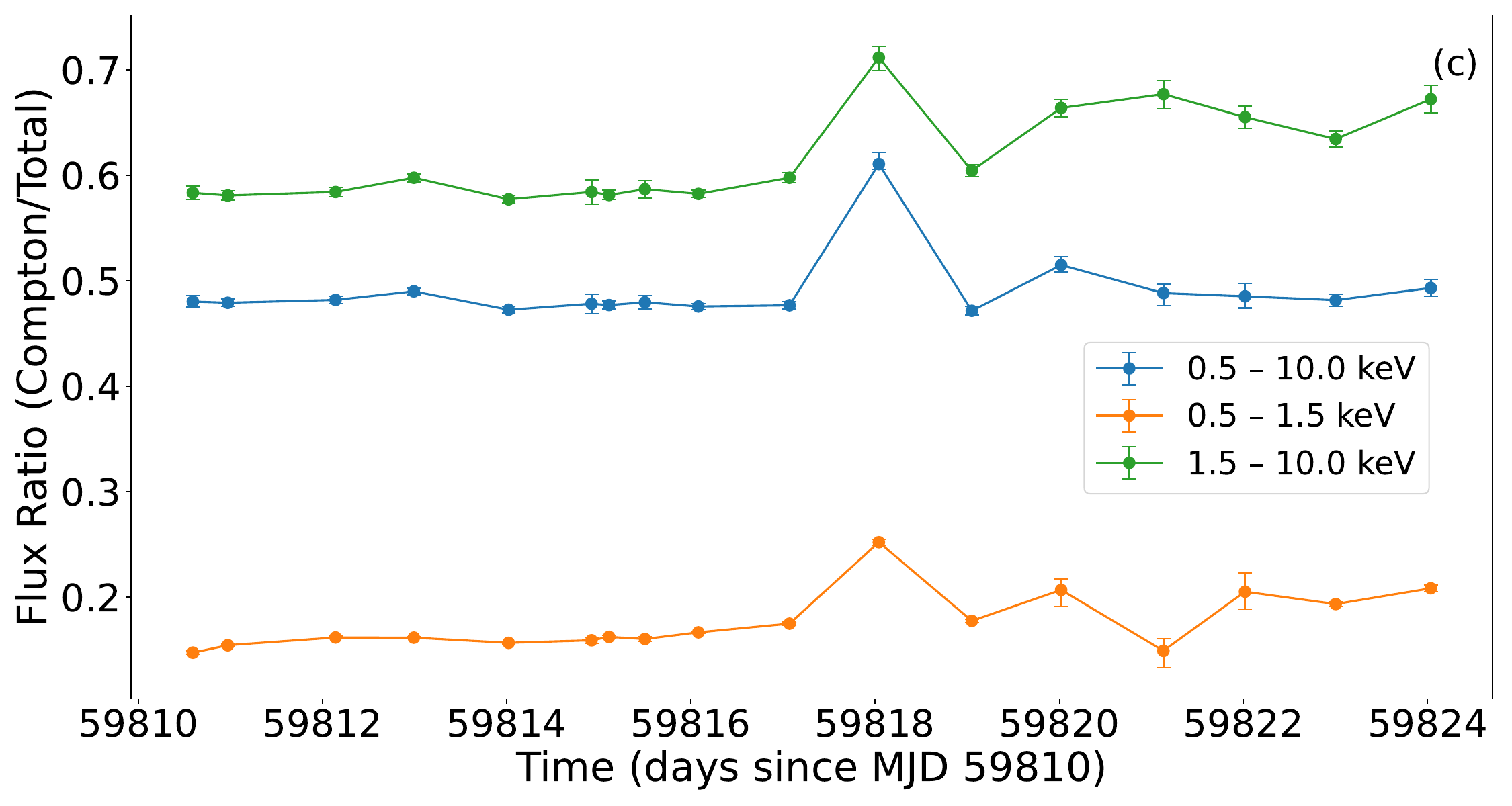}
\includegraphics[width=0.45\columnwidth]{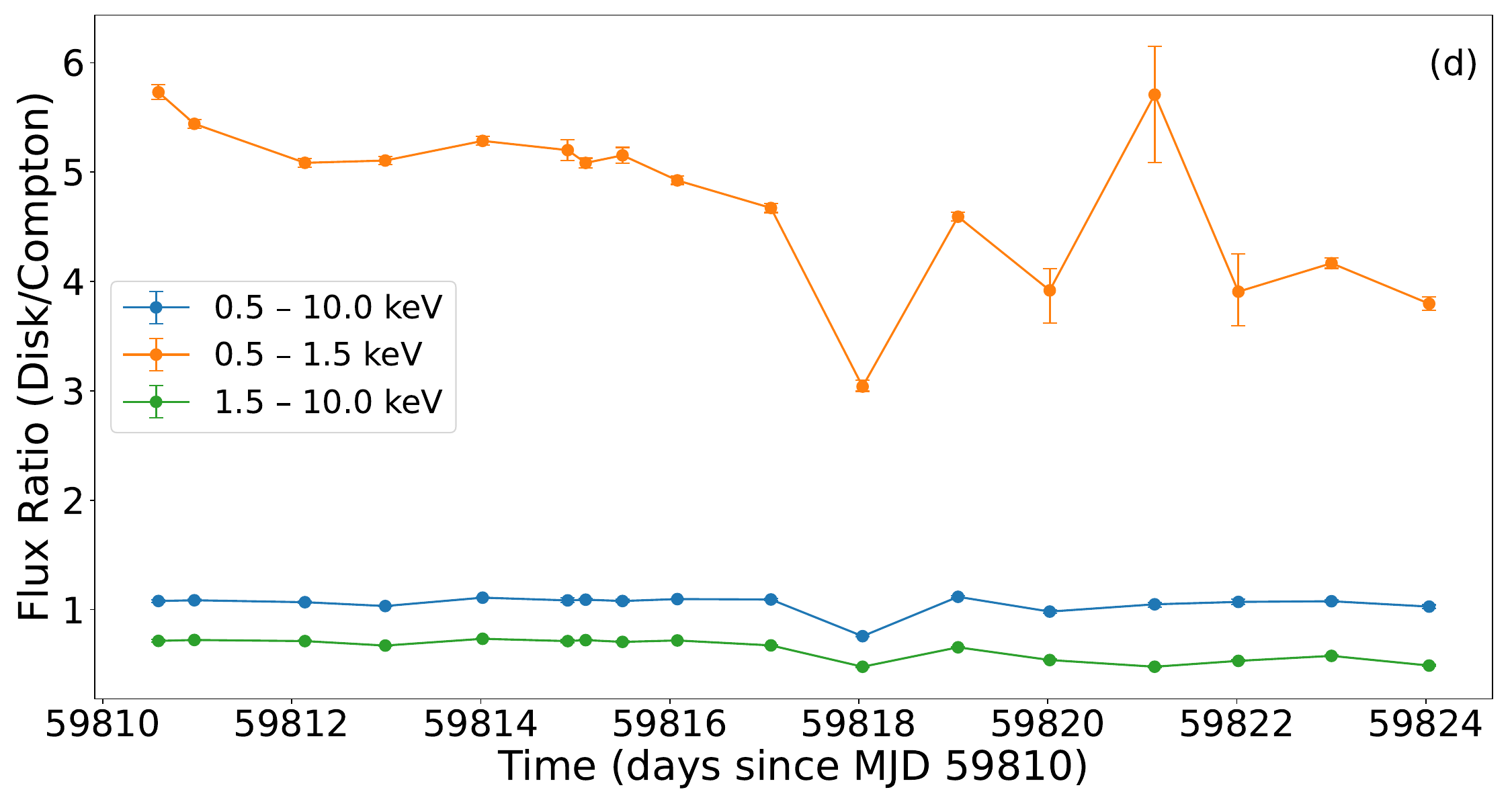}
\caption{The plots show the evolution of unabsorbed flux in different energy bands for the source SAX J$1808.4-3658$ observed during its 2022 outburst. Panel (a): The evolution of total flux, disk flux and flux from the corona, in the energy range $0.5-10$\,keV. Panel (b): The ratio of disk flux to the total flux in three energy bands. Panel (c): The ratio of corona flux to the total flux in three energy bands. Panel (d): The ratio of disk flux to corona flux in the three energy bands.}
\label{fig:appendix_plots}
\end{figure*}

\bibliography{ref}
\bibliographystyle{aasjournal}

\end{document}